\documentclass[10pt,aps,pre,amsmath,amsfonts,amssymb,floatfix,longbibliography,notitlepage,twocolumn]{revtex4}
\usepackage{mathrsfs} \usepackage{graphicx,color} \usepackage{bbm} \usepackage{multirow}
\usepackage{bm}

\let\a=\alpha \let\b=\beta \let\g=\gamma \let\d=\delta
   
\let\l=\lambda    \let\p=\pi
\let\s=\sigma \let\t=\tau \let\f=\varphi \let\ph=\varphi
   
\let\D=\Delta \let\L=\Lambda  
   
 \let\r=\rho \let\th=\theta \let\io=\infty

\def\NN{{\cal N}} 
  
 \def\SS{{\cal S}}
  
\def\ZZ{{\cal Z}}

\def\ol{\overline}

\newcommand{\br}{\mathbf{r}}
\newcommand{\bk}{\mathbf{k}}

\def\to{\rightarrow} \def\la{\left\langle} \def\ra{\right\rangle}

\newcommand{\beq}{\begin{equation}} \newcommand{\eeq}{\end{equation}}

\begin{document}

%% For titles, only capitalize the first letter
\title{Hopping and the Stokes--Einstein relation breakdown in simple glass formers}

%% Enter authors via the \author command.
%% Use \affil to define affiliations.
%% (Leave no spaces between author name and \affil command)

%% \author{<author name>
%% \affil{<number>}{<Institution>}} One number for each institution.
%% The same number should be used for authors that
%% are affiliated with the same institution, after the first time
%% only the number is needed, ie, \affil{number}{text}, \affil{number}{}
%% Then, before last author ...
%% \and
%% \author{<author name>
%% \affil{<number>}{}}

%% For example, assuming Garcia and Sonnery are both affiliated with
%% Universidad de Murcia:
%% \author{Roberta Graff\affil{1}{University of Cambridge, Cambridge,
%% United Kingdom},
%% Javier de Ruiz Garcia\affil{2}{Universidad de Murcia, Bioquimica y Biologia
%% Molecular, Murcia, Spain}, \and Franklin Sonnery\affil{2}{}}

\author{Patrick Charbonneau}
\affiliation{Departments of Chemistry and Physics, Duke University, Durham,
North Carolina 27708, USA}
\author{Yuliang Jin}
\affiliation{Departments of Chemistry and Physics, Duke University, Durham,
North Carolina 27708, USA}
\affiliation{Dipartimento di Fisica,
Sapienza Universit\'a di Roma, INFN, Sezione di Roma I, IPFC -- CNR, P.le A. Moro 2, I-00185 Roma, Italy}
\author{Giorgio Parisi}
\affiliation{Dipartimento di Fisica,
Sapienza Universit\`a di Roma, INFN, Sezione di Roma I, IPFC -- CNR, P.le A. Moro 2, I-00185 Roma, Italy}
\author{Francesco Zamponi}
\affiliation{LPT,
\'Ecole Normale Sup\'erieure, UMR 8549 CNRS, 24 Rue Lhomond, 75005 France}

%\contributor{Submitted to Proceedings of the National Academy of Sciences
%of the United States of America}

%%%%%%%%%%%%%%%%%%%%%%%%%%%%%%%%%%%%%%%%%%%%%%%%%%%%%%%%%%%%%%%%
%\begin{article}

\begin{abstract} %%217 words < 250 words max
One of the most actively debated issues in the study of the glass transition is whether a mean-field description is a reasonable starting point
for understanding experimental glass formers.
Although the mean-field theory of the glass transition --
like that of other statistical systems -- is exact when the spatial dimension $d\to\io$,
the evolution of systems properties with $d$ may not be smooth.
Finite-dimensional effects could dramatically
change what happens in physical dimensions, $d=2,3$. For standard phase transitions finite-dimensional effects are typically captured by
renormalization group methods, but for glasses the corrections are much more subtle and only partially understood.
Here, we investigate hopping between localized
cages formed by neighboring particles in a model that allows to cleanly isolate that effect.
By bringing together results from replica theory, cavity reconstruction, void percolation, and molecular dynamics, we obtain insights into
how hopping induces a breakdown of the Stokes--Einstein relation and modifies the mean-field scenario in experimental systems.
Although hopping is found to supersede the dynamical glass transition, it nonetheless leaves a sizable part of the critical regime untouched.
By providing a constructive framework for identifying and quantifying the role of hopping,
we thus take an important step towards describing dynamic facilitation in the framework of the mean-field theory of glasses.
\end{abstract}

%% When adding keywords, separate each term with a straight line: |
\maketitle

%% Optional for entering abbreviations, separate the abbreviation from
%% its definition with a comma, separate each pair with a semicolon:
%% for example:
%% \abbreviations{SAM, self-assembled monolayer; OTS,
%% octadecyltrichlorosilane}

% \abbreviations{}

%% The first letter of the article should be drop cap: \dropcap{}
%\dropcap{I}n this article we study the evolution of ''almost-sharp'' fronts

%% Enter the text of your article beginning here and ending before
%% \begin{acknowledgements}
%% Section head commands for your reference:
%% \section{}
%% \subsection{}
%% \subsubsection{}

\noindent \noindent\rule{8cm}{0.4pt}
%------------------------------------------------------------------------------------------------------------------------------------------------------
\\
{\bf Significance} %%%%92 words <120 words max
Like crystals, glasses are rigid because of the {\it self-caging} of their constituent particles.
The key difference is that 
%{\color{red} 
crystal formation is a sharp first order 
phase transition at which cages form abruptly and remain stable, while glass formation entails the progressive emergence of cages. This 
%}
loose caging complicates the description of the glass transition. In particular, an important transport mechanism in this regime, hopping, has thus far been difficult to characterize.
Here we develop a completely microscopic description of hopping, which allows us to clearly assess its impact on transport anomalies, such as the breakdown of the Stokes--Einstein relation.

%The scientific landscape of the glass transition is notoriously contentious. After decades of study, the theoretical description of this everyday phenomenon has remains fraught with fundamental conflicts about the physical origin of the dynamical arrest. In this work, we unify two descriptions that are often pitted against one another by developing mathematical and physical tools to disentangle their respective contributions. The resulting insights will likely bring a more ecumenical future.

%\vskip5pt

%{\bf Other proposal --}
%Rigidity of glasses comes from the {\it caging} of particles by neighboring ones, like in crystals.
%At the liquid-crystal transition, cages form abruptly and they remain stable throughout the crystal phase,
%hence hopping of particles between different cages remains highly suppressed and does not contribute
%to transport. The glass transition is instead a continuous transition where caging emerges progressively
%in the liquid. In the transition region, cages are loose enough that hopping is an important mechanism for transport, but its
%treatment remained so far mostly phenomenological.
%In this paper we develop for the first time a completely microscopic description of the hopping process, and we are able
%to assess its impact on transport, e.g. the violation of the Stokes-Einstein relation.

\noindent\rule{8cm}{0.4pt}
%------------------------------------------------------------------------------------------------------------------------------------------------------

\paragraph*{Introduction -}
%Amorphous materials such as grains, foams, and glasses are encountered daily. Controlling their properties makes for better pharmaceuticals, cosmetics, and various other household products, and is also crucial for engineering smart materials, improving industrial processes, and understanding avalanches. Yet a theoretical description of these materials at the microscopic level is still missing, hindering progress, which has so far been largely empirical. Developing a theoretical framework for describing these materials is extremely challenging. Conventional theoretical paradigms based on perturbative expansions around the low-density, ideal gas limit (for moderately dense gases and liquids), or on harmonic expansions around an ideal lattice (for crystals) fail badly. Because dense amorphous materials interact strongly, low-density expansions are unreliable, while harmonic expansions lack reference equilibrium particle positions. These fundamental difficulties ought to be surmounted in order to describe the ubiquitous non-equilibrium process often dubbed the \emph{glass} transition.

Glasses are amorphous materials whose rigidity emerges from the mutual caging of their constituent particles -- be they atoms, molecules, colloids, grains, or cells.
Although glasses are ubiquitous, the microscopic description of their formation, rheology, and other dynamical features is still far from satisfying. Developing a more complete theoretical framework would not only resolve epistemological wrangles~\cite{BBBCS11}, but also improve our material control and design capabilities. Yet such a research program remains fraught with challenges. Conventional paradigms based on perturbative expansions around the low-density, ideal gas limit (for moderately dense gases and liquids), or on harmonic expansions around an ideal lattice (for crystals) fail badly. Because dense amorphous materials interact strongly, low-density expansions are unreliable, while harmonic expansions lack reference equilibrium particle positions. These fundamental difficulties must somehow be surmounted in order to describe the dynamical processes at play in glass formation.

A celebrated strategy for studying phase transitions 
 is to consider first their mean-field description,
which becomes exact when the spatial dimension $d$ of the system goes to infinity~\cite{Wi80}, before including corrections to this description.
In that spirit, we open with the $d\rightarrow\infty$ ``ideal'' random first-order transition (iRFOT) scenario, 
%{\color{red} 
which, based on the analysis of simple models,
%} 
brings together static-~\cite{FP97,MP00,BB04} and dynamics-based (mode-coupling)~\cite{Go09} results for glass formation
(see, e.g.,
%{\color{red}
~\cite{BB11,PZ10}
%} %~\cite{CC05,BB11,PZ10} 
for reviews)~\cite{KW87,PZ10,KPZ12,CKPUZ14}. %KPUZ13,CKPUZ13,CKPUZ14}.
In iRFOT, an infinitely slowly cooled simple liquid (or compressed hard sphere fluid) becomes infinitely viscous, i.e., forms a glass in which 
 particles are completely caged, at the (critical) dynamical transition temperature $T_{\mathrm{d}}$ (or packing fraction $\varphi_{\mathrm{d}}$).
%According to mode-coupling theory~\cite{Go09},  
Upon approaching this transition,
caging makes the diffusivity $D$ vanish as a power-law $D \sim ( T - T_{\rm d} )^\g$,
and the viscosity diverge as $\eta\sim (T-T_\mathrm{d})^{-\gamma}$. Hence, in the critical regime one expects the 
Stokes--Einstein relation (SER) between transport coefficients, $D \sim \eta^{-1}$, to hold.
In short, the $d\rightarrow\infty$ scenario is characterized by (i)~a sharp dynamical glass transition associated with perfect caging,
(ii)~a power-law divergence of $\eta$, and (iii)~the SER being obeyed.

As observed in Ref.~\cite{KTW89},
the phenomenology of finite-dimensional systems is, however, quite different from the iRFOT scenario. In particular, it
does not recapitulate elementary experimental observations, such as
Vogel-Tammann-Fulcher (VTF) viscosity scaling in fragile glasses, 
$\eta\sim e^{B_{\rm VTF}/(T-T_0)}$ ($B_{\rm VTF}$ and $T_0$ are phenomenological constants),
and breakdown of the SER, $D\sim\eta^{-1+\omega}$ (phenomenologically $\omega>0$)~\cite{DS01,KSD06,ER09,CCJPZ13}.
As a result, the relevance of the iRFOT picture for experimental systems remains the object of lively debates.

%{\color{red} 
Part of the difficulty of clarifying the situation  in finite $d$, where the iRFOT description is only approximate and the dynamical transition is but a  crossover, lies in the shear number of different contributions one has to take into account.
%}
 From a purely field-theoretic point of view, one has to include finite-dimensional corrections to
critical fluctuations. A Ginzburg criterion gives $d_u=8$ as the upper critical dimension for the dynamical transition~\cite{BB07,FPRR11,FJPUZ12,BCTT14}, and hence for $d < d_u$
critical fluctuations renormalize the power-law scaling exponents. In principle, these corrections could be captured by a perturbative $d_u-d$ expansion, and phenomenological arguments
along this direction indicate that they could also induce a SER breakdown~\cite{BB07}. A number of non-perturbative processes in $1/d$ must additionally be considered. (i)~In the
iRFOT picture, caging is perfect, hence in the glass phase each particle is forever confined to a finite region of space delimited by its
neighbors~\cite{Go09}. However, it has been theoretically proven~\cite{Os98} and experimentally observed~\cite{CDB09} that in 
%{\color{red}
low-dimensional systems the diffusivity is never strictly zero.
%}
 Single
particles can indeed hop between neighboring cages~\cite{SS03,Sc05,CBK07,MS13}, and the free space they leave behind can {\it facilitate} the hopping of
neighboring particles. Facilitation can thus result in cooperative hopping and avalanche formation~\cite{GC03,CWKDBHR10,KHGGC11}. (ii)~For some glass formers, activated crystal
nucleation cannot be neglected and interferes with the dynamical arrest, leading to a glass composed of microscopic geometrically frustrated crystal domains~\cite{TKSW10}. (iii)~In the
iRFOT scenario, the dynamical arrest is related to the emergence of a huge number of distinct metastable glass states whose lifetime is infinite. %A system confined to one of these states remains within it. 
In finite dimensions, however, a complex glass-glass nucleation process gives a finite lifetime to these metastable states~\cite{KTW89,XW01a,BB04}. The dynamics of glass-forming liquids is then profoundly affected. Including glass-glass nucleation into iRFOT leads to the complete 
RFOT scenario~\cite{KTW89}, in which
the mean-field dynamical glass transition becomes but a crossover~\cite{KTW89}, 
and both VTF scaling and facilitation are recovered~\cite{BBW08,WL12}.

Because
the treatment of these different processes has thus far been mostly qualitative, their relative importance cannot be easily evaluated.
A controlled first-principle, {\it quantitative} treatment is for the moment limited to the exact solution for 
$d\to\io$~\cite{KPZ12,KPUZ13,CKPUZ13,CKPUZ14}. Its approximate extension to finite 
$d$~\cite{MP09,PZ10,Go09} completely ignores the non-perturbative effects mentioned above.
This approach therefore cannot, on its own, cleanly disentangle the various corrections.
Systematic studies of glass formation as a function of $d$ have encouragingly shown that these corrections are limited,
even down to $d=3$~\cite{SDST06,ER09,CIPZ11,CCT13,SKDS13,CCJPZ13}, provided length and time scales are not too large,
as is typical of numerical simulations and experiments with
colloids and grains. In particular, with increasing $d$ the distribution of particle displacements (the self-van Hove function) %$G_s(r,t)$) 
loses its second peak associated with hopping~\cite{CCJPZ13}, the critical power-law regimes lengthen~\cite{CIPZ12}, and the SER breakdown weakens~\cite{ER09,CCJPZ13,SKDS13},
%{\color{red}
 which motivates investigating corrections to iRFOT in a controlled way.
%}

Here we develop a way to isolate the simplest of these corrections, i.e., hopping,
by studying a finite-dimensional mean-field model. Through the use of the cavity reconstruction methodology developed in the context of spin glass and information theory
%{\color{red}
~\cite{MM06},
%}
 %~\cite{MM09},
we carefully describe caging using self-consistent equations that can be solved numerically. We can thus compute the cage width
distribution and isolate hopping processes. % to study their properties.
Our results provide an unprecedentedly clear view of the impact of hopping on the dynamical transition
and on the SER breakdown in simple glass formers.

\paragraph*{MK Model -}
We consider 
%{\color{red} 
the infinite-range variant of
%} 
the hard sphere (HS)-based model proposed by Mari and Kurchan (MK) for simple structural glass formers~\cite{MKK08,MK11,MPTZ11} (see SI Sec.\;IA for details). 
The key feature of the MK model is that, even though each sphere has the same diameter $\sigma$, pairs of spheres interact via an additional constant shift that is randomly-selected over the full system volume.
This explicit quenched disorder eliminates the possibility of a crystal state, 
suppresses coherent activated barrier crossing that leads to glass-glass nucleation~\cite{MK11}, and diminishes the possibility of facilitated hopping (as we discuss below).
%In addition, the model's critical properties correspond to those of the iRFOT description~\cite{MK11}. 
Yet 
%{\color{red} 
at finite densities
%} 
the number of neighbors that interact with a given particle is finite 
and therefore finite-dimensional corrections related to hopping remain, in principle, possible.

MK liquids have a trivial structure.
Even in the dense and strongly interacting regime, the pair correlation in the liquid phase is simply $g_2(r)=\theta(r-\sigma)$ (where $\theta(x)$ is the Heaviside step function),
because particles are randomly displaced in space. In addition,
even if both particles $i$ and $k$ are nearby particle $j$ they need not be close neighbors, hence
all higher-order structural correlations are perfectly factorizable. Because only two-body correlations contribute, the virial series can be truncated
at the second virial coefficient~\cite{MK11}, hence
the equation of state for pressure is trivially $\beta P/\rho=1+B_2 \rho$, where $B_2=V_d(1) \s^d/2$ is the second-virial coefficient
for $d$-dimensional hard spheres, $V_d(R)$ is the volume of a $d$-dimensional ball of radius $R$,
$\rho$ is the number density (the packing fraction $\f = \r V_d (\s/2)$),
and the inverse temperature $\beta$ is set to unity~\cite{MKK08,MK11,MPTZ11} (see SI Sec.\;IA).
Note that these structural features hold for the liquid phase of the MK model in all $d$, {\it and} for standard HS liquids in the limit $d\to\io$~\cite{FP99,PZ10}. The MK model therefore coincides with standard HS in that limit. For a given finite $d$, however,
MK liquids are structurally more similar to their $d\to\io$ counterparts than HS liquids are. One thus sidesteps having to take into account
the non-trivial structure of $g_2(r)$, which muddles the description of standard finite-dimensional HS~\cite{PZ10}.

For the MK model, one can easily construct \emph{equilibrated} liquid configurations at all $\f$, even for 
%{\color{red} 
$\f>\f_{\mathrm d}$
%}
(For standard HS, by contrast, prohibitively long molecular dynamics (MD) simulations are necessary in this regime.)
This dramatic speedup is accomplished by adapting the
{\it planting} technique developed in the context of information theory~\cite{KZ09} (see SI Sec.\;IB).
It is thus possible to study MK liquids arbitrarily close to, both above and below, the dynamical glass transition at $\varphi_\mathrm{d}$.
A systematic study of caging beyond $\varphi_{\mathrm{d}}$ is also possible thanks to the cavity reconstruction formalism,
a method adapted from the statistical physics of random networks~\cite{MM06}.

\begin{figure}[t]
\includegraphics[width =\columnwidth ]{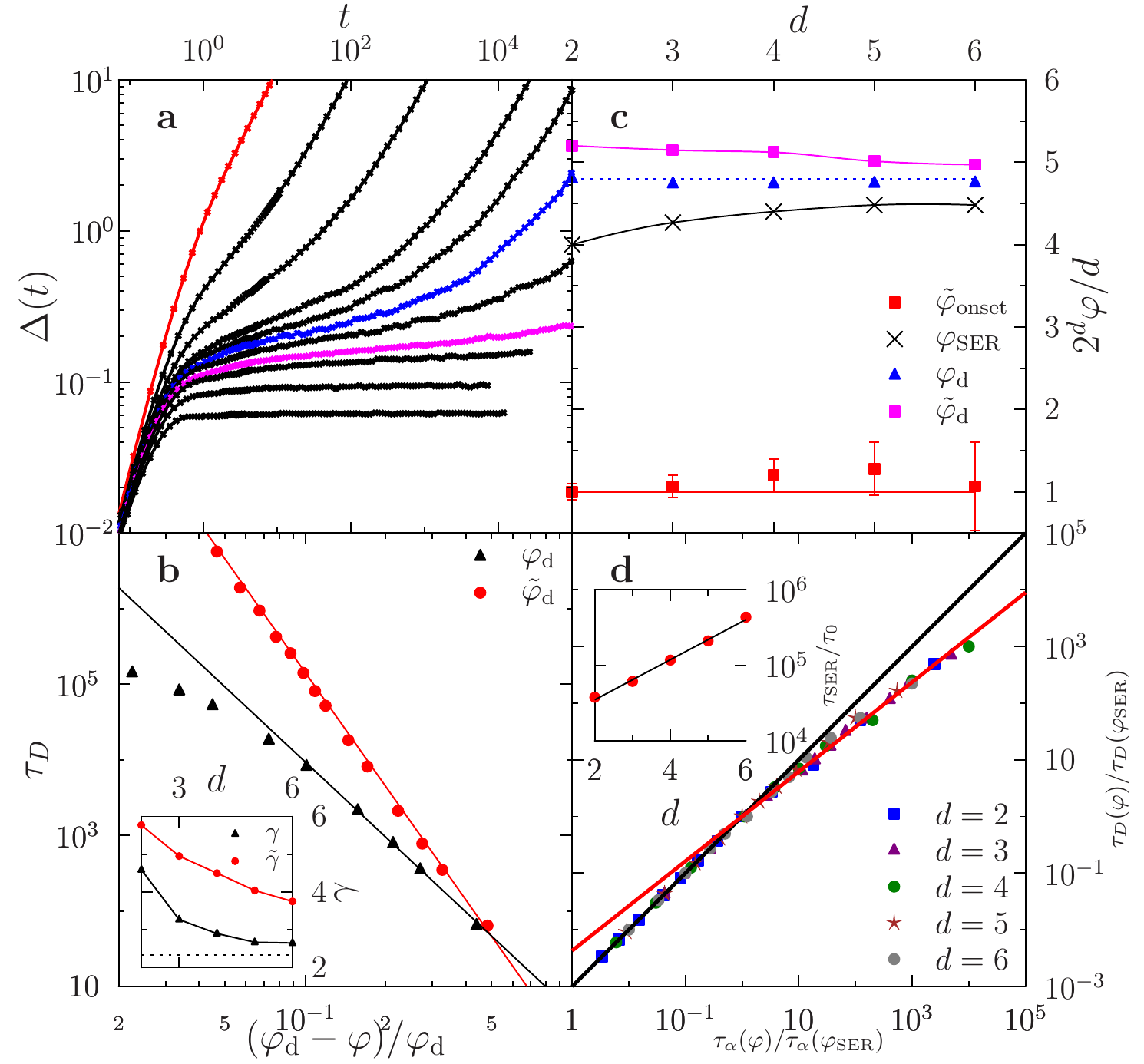}
\caption{(a) MSD of the MK model in $d=3$ for $\varphi=$0.40, 1.00, 1.40, 1.65, 1.72, 1.78, 1.84, 1.93, 2.00, 2.20, and 2.50, from top to bottom. The onset of caging $\tilde{\varphi}_{\mathrm{onset}}$ (red), the theoretical dynamical transition $\varphi_{\mathrm{d}}$ (blue), and its dynamical estimate $\tilde{\varphi}_{\mathrm{d}}$ (magenta) 
%{\color{red} 
are highlighted.
%}
 Note that at $\tilde{\varphi}_{\mathrm{d}}$ and beyond a steady drift of the MSD plateau can be detected. (b) Power-law scaling in $d$=3 of the characteristic time $\tau_D$ determined by fitting   $\tilde{\varphi}_{d}$=1.93 and $\tilde{\gamma}$=4.95 and by using the idealized mean-field result $\varphi_\mathrm{d}$=1.78
and by fitting $\gamma$=3.27. (inset) Dimensional evolution of $\gamma$ and $\tilde{\gamma}$. The dashed line indicates the $d=\infty$ result $\gamma$=2.33786~\cite{KPUZ13}. Solid lines are guides for the eye. (c) The dimensional scaling of $\tilde{\varphi}_\mathrm{d}$, $\varphi_\mathrm{d}$, and $\varphi_{\mathrm{SER}}$ converges as $d$ increases, while the onset of caging at $\tilde{\varphi}_{\mathrm{onset}}$ remains clearly distinct. 
%{\color{red} 
The dashed line is the  replica result $\varphi_{\mathrm d}=4.8d2^{-d}$~\cite{PZ10,KPZ12}.
%}
Solid lines are guides for the eye. (d) Dimensional rescaling of the SER (black) and SER breakdown (red) regimes for the MK model with $\omega$=0.22.  
%{\color{red} 
(inset) The ratio $\tau_{\rm SER}/\tau_0$ grows exponentially with $d$ (solid line), where $\tau_{\rm SER} = \tau_D (\ph_{\rm SER})$ and $\tau_0$ is the microscopic time, i.e., the characteristic time for the decay of the velocity autocorrelation function~\cite{Ikeda2013} (see SI Sec.\;IC2 for details).
%}
\label{fig:MSD}}
\end{figure}

\paragraph*{Caging - } The MK model dynamics is studied by event-driven MD simulations of planted initial configurations 
%{\color{red} 
with $N=4000$ particles (see SI Sec.\;IB for details)~\cite{SDST06,CIPZ11}. The mean square displacement (MSD)  $\Delta(t)=\langle\sum_{i=1}^N [\mathbf{r}_i(t)-\mathbf{r}_i(0) ]^2\rangle/N$
is determined from time evolution of the particle positions $\mathbf{r}_i(t)$.
At short times, before any collision occurs, ballistic motion gives $\D(t) =d t^2$;
at long times, diffusive motion gives $\D(t) \sim 2 d D t$.
%}
From $\tilde{\varphi}_{\mathrm{onset}}$ onwards, the ballistic and the diffusive regimes are separated by an intermediate caging regime where
$\D(t) \approx \bar{\Delta}$ is approximately constant, first appearing as an inflection point and then as a full-fledged plateau (see SI Sec.\;IC for definition).
Simply put, after a few collisions with its neighbors, a particle becomes confined to a small region of space of linear size $\sqrt{\bar{\Delta}}$,
from which it can only escape, and henceforward diffuse, after a very large number of collisions.

In the $d\rightarrow\infty$ iRFOT scenario, a sharp
dynamical transition occurs at $\f_{\rm d}$~\cite{Go09,PZ10,KPZ12}, beyond which complete caging results in an infinitely-long plateau and in the
disappearance of the diffusive regime.
In finite-dimensional systems, one can use an approximate theory based on a Gaussian assumption for the cage shape, to obtain a prediction for $\f_{\rm d}$
and $\Delta$~\cite{PZ10,MPTZ11} (see SI Sec.\;IIA). One can also estimate $\tilde{\varphi}_{\mathrm{d}}$ from the simulation results by fitting the diffusivity using the mean-field critical form
$D\sim(\varphi-\tilde{\varphi}_{\mathrm{d}})^{\tilde{\gamma}}$, and $\bar{\Delta} = \lim_{t\rightarrow\infty}\Delta(t)$  beyond $\tilde{\varphi}_{\mathrm{d}}$
(Fig.~\ref{fig:MSD}a).  As expected from the suppression of various finite $d$
corrections, the critical power-law regime is much longer for the MK model than for standard finite-dimensional HS (Fig.~\ref{fig:MSD}b)~\cite{MK11}.
Marked qualitative discrepancies from the iRFOT predictions are nonetheless observed.
(i)~Numerical estimates for $\tilde{\varphi}_\mathrm{d}$ systematically deviate from the approximate Gaussian result for $\f_d$ (Fig.~\ref{fig:MSD}c), 
even though the two quantities grow closer with dimension.
(ii)~The diffusion time
%$\tau_D= \sigma^2/D$ \added{why this prefactor? I think we can say $\tau_D \sim 1/D$.}
$\tau_D= \sigma^2/D$
and the structural relaxation time $\tau_\alpha\propto\eta$ (see SI Sec.\;IC for definitions and a discussion of this point)
follow the SER, $\tau_D \propto \tau_\alpha$, from $\tilde{\varphi}_{\mathrm{onset}}$ to $\varphi_{\mathrm{SER}}<\tilde{\varphi}_{\mathrm{d}}$, but then the SER breaks down, $\tau_D\propto\tau_\alpha^{1-\omega}$ with $\omega\approx0.22$, in all $d$ (Fig.~\ref{fig:MSD}d). 
%{\color{red} 
With increasing $d$, however, the timescale for this crossover,  $\tau_D(\varphi_{\mathrm{SER}})$,  also increases (Fig.~\ref{fig:MSD}d),
%}
and thus $\varphi_{\mathrm{SER}}$ grows closer to $\varphi_{\mathrm{d}}$ and $\tilde{\varphi}_{\mathrm{d}}$  (Fig.~\ref{fig:MSD}c).
(iii)~Even above $\tilde{\varphi}_{\mathrm{d}}$, a steady drift of the MSD plateau can be detected (Fig.~\ref{fig:MSD}a), but the magnitude of this effect diminishes with increasing $d$.

\begin{figure}[t]
\includegraphics[width =\columnwidth]{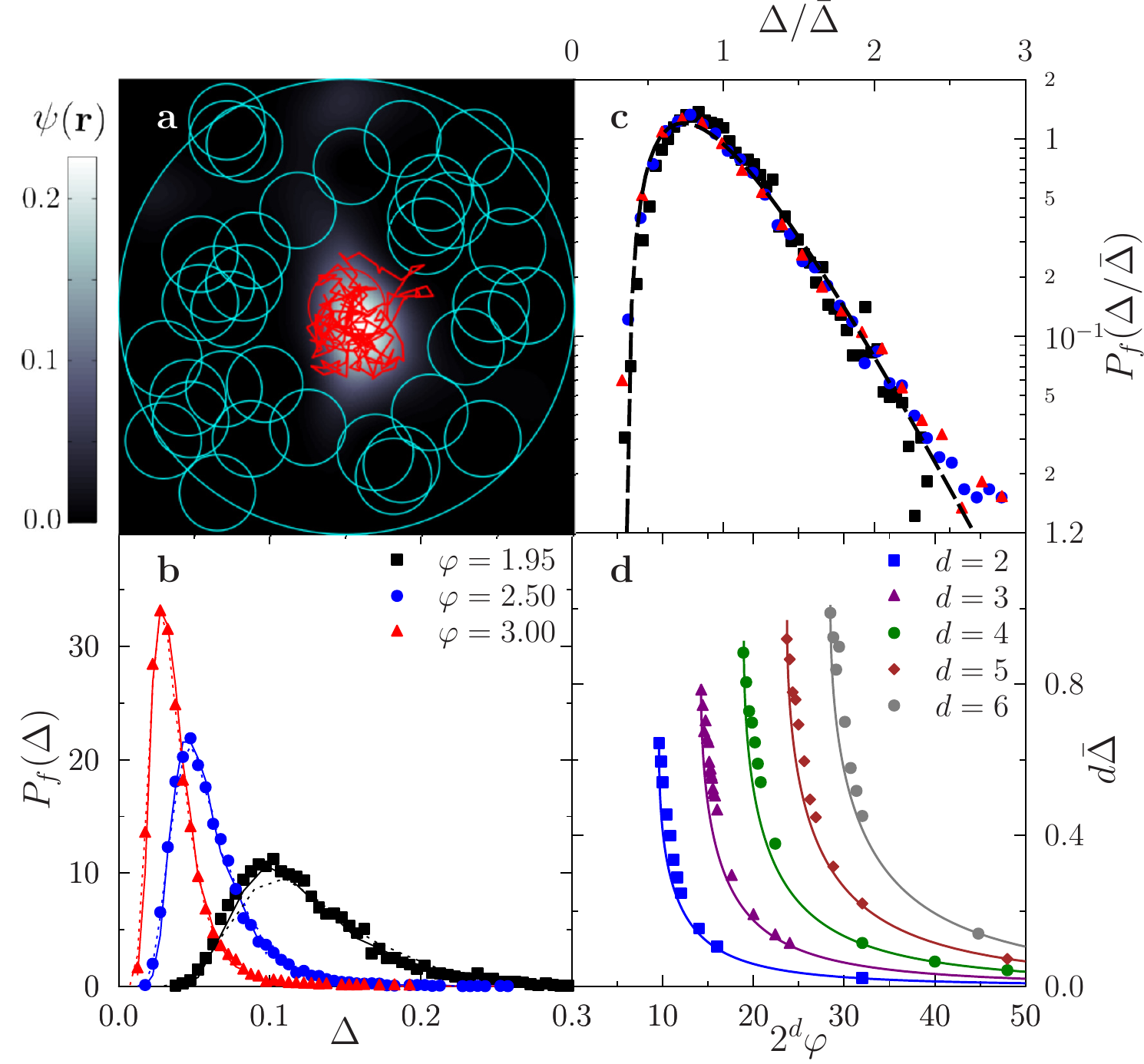}
\caption{(a) Illustration of a cavity reconstruction in $d$=2 for a perfectly caged particle at the center. Neighboring particles at their equilibrium positions (circles) provide an effective field $\psi(\mathbf{r})$ that cages the trajectory of the central particle (red line). (b) Examples of $P_f(\Delta)$ in $d$=3 from the cavity reconstruction formalism for Gaussian (straight lines) and ball (dashed lines) cage shapes compared with MD results (symbols). (c) Rescaled $P_f(\Delta)$ superimposed with a log normal distribution (dashed line). (d) Density evolution of $\bar{\Delta}$ measured from MD simulations (points) superimposed on the theoretical predictions of Refs.~\cite{PZ10,MPTZ11} (lines).\label{fig:PA}}
\end{figure}

In order to clarify the physical origin of the above discrepancies, we first determine whether the mismatch between $\tilde{\varphi}_{\mathrm{d}}$ and $\varphi_{\mathrm{d}}$
is due to the hypothesis made in computing the latter, i.e., that all the cages have a Gaussian shape of a fixed diameter $\bar{\Delta}$,
by using the cavity reconstruction formalism to relax both assumptions~\cite{MM06}. %,MM09}.
Above $\tilde{\varphi}_{\mathrm{d}}$, we can build the equilibrated neighborhood of particle $i$ in order to self-consistently determine
the overall cage size and/or shape distribution $P_f(\Delta)$ (see SI Sec.\;IIA for details). The process involves placing Poisson-distributed neighbors
$j$ that are randomly assigned a cage size $\Delta_j$ from a prior guess of $\tilde{P}_f(\Delta)$, with a fixed
function shape $f_{A_j}(\mathbf{r})$ (a Gaussian or a ball function, for instance).
Averaging over the vibrational relaxation of each neighboring particle gives the cavity field
$\psi(\mathbf{r})$  % \propto \sum_j q_{A_j}(\mathbf{r})$
felt by particle $i$, which is the probability density of the particle being at position $\mathbf{r}$
(Fig~\ref{fig:PA}a). The existence of a cage centered around $i$ is guaranteed by the cavity reconstruction procedure.
The variance $\langle \delta \mathbf{r}^2 \rangle = \langle \mathbf{r}^2 \rangle- \langle \mathbf{r} \rangle^2$ associated with the evolution of particle $i$ within this cage, which can be computed through simple Monte Carlo
sampling, provides the posterior caging radius $\Delta_i$. Sufficient repeats of this determination provides a new estimate of
$\tilde{P}(\Delta)$, and iterating the overall procedure
eventually converges to a fixed point distribution $P_f(\Delta)$. % after sufficiently many iterations.
We find that both Gaussian and ball caging functions give the same size distribution $P_f(\Delta)$ (Fig~\ref{fig:PA}b),
and that $P_f(\Delta)$ is reasonably well approximated by a gamma distribution for all $\varphi>\varphi_{\mathrm{d}}$ (Fig~\ref{fig:PA}c).
The average cage size $\bar{\Delta}$ also quantitatively agrees with the analytical prediction of Refs.~\cite{PZ10,MPTZ11} (Fig~\ref{fig:PA}d),
including its characteristic square-root singularity upon approaching $\varphi_{\mathrm{d}}$, i.e.,
$\bar{\Delta}(\varphi_\mathrm{d})-\bar{\Delta}(\varphi)\sim \sqrt{\varphi-\varphi_{\mathrm{d}}}$.
%{\color{blue} 
Therefore, the theoretical prediction of $\bar{\Delta}$ and $\varphi_{\rm d}$ is fairly insensitive to the caging form and the second (or higher) moments of the cage size distribution, as well as to the method we choose (see SI Sec.~IIA).
%}
%MK caging is therefore fairly insensitive to the caging form as well as to the second and higher moments of the cage size distribution.

\begin{figure}[t]
\includegraphics[width =\columnwidth]{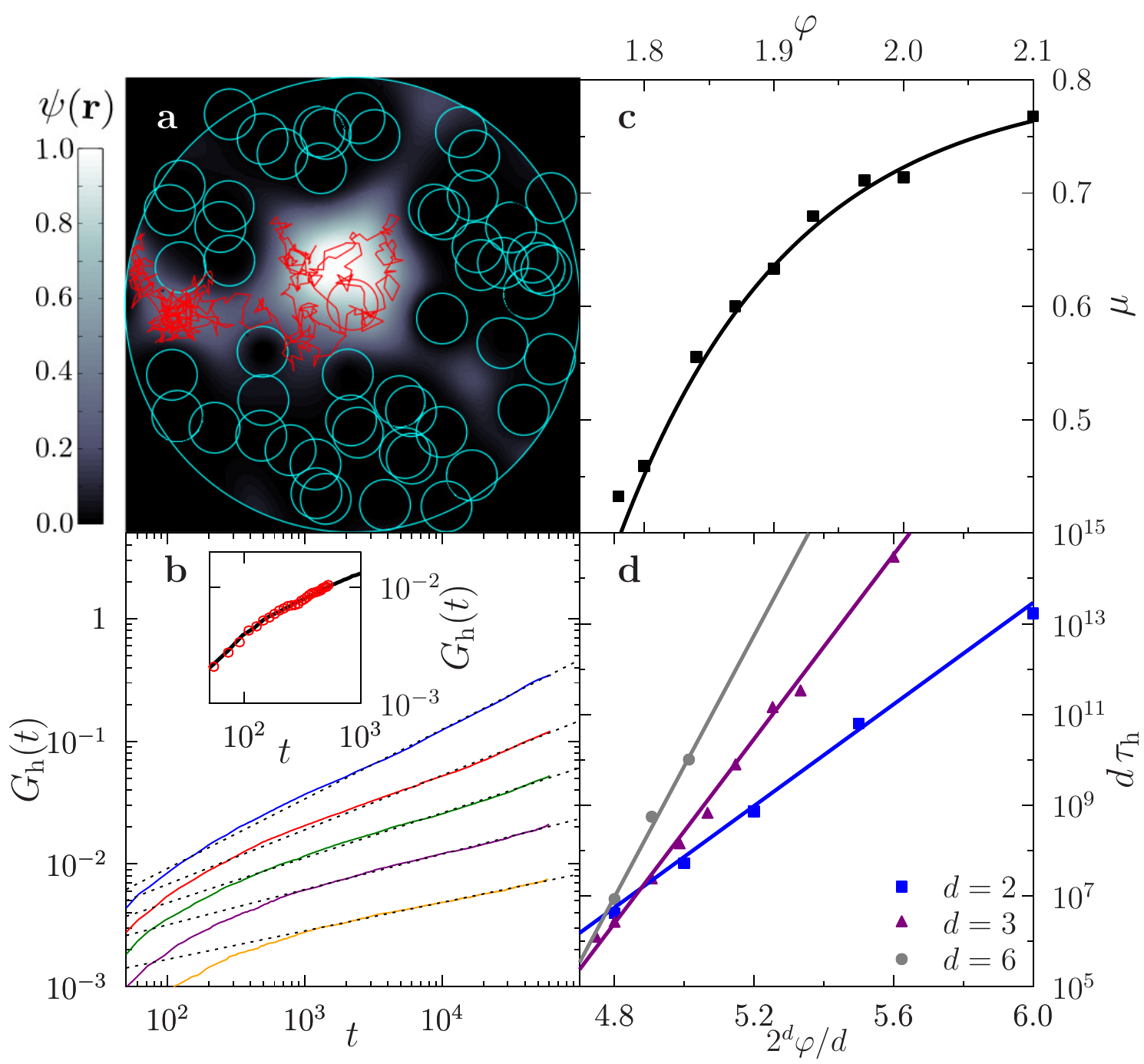}
\caption{(a) Illustration of a cavity reconstruction in $d=2$ for a hopping particle. In this case the neighboring particles allow the central particle to hop to other cages (red line).
%(b) Typical temporal and spatial (along the $x$ coordinate for a system of side $L$) distribution of hopping events in a given MD trajectory of $N$ particles in $d$=3 at $\varphi$=1.85.
(b) Cumulative time probability distribution of hopping events $G_{\mathrm{h}}(t)$ for $d=3$ systems at densities (from top to bottom) $\ph =1.78, 1.84, 1.90, 1.97$, and 2.10, along with the power-law scaling form (dashed line). (inset) Single-particle hopping from the cavity reconstruction (circles) overlays with the MD simulations at short times ($\varphi = 1.90$). Phenomenological scaling parameters
(c) $\mu$ and (d) $\tau_{\rm h}$ for the probability distribution of hopping events. 
%{\color{red} 
Solid lines are a guide for the eye for $\mu$ and exponential fits for $\tau_{\rm h}$.}\label{fig:hopping}
%}
\end{figure}

It follows that deviations from the $d\to\infty$ scenario ought to be ascribed to an imperfect caging above $\varphi_{\mathrm{d}}$ in finite-dimensional systems.
Microscopically, these imperfections correspond to particles  trapped for a finite time before escaping to another cage through a narrow passage (Fig.~\ref{fig:hopping}a).  Because the above calculations solely consider single-cage forms, a fixed-point distribution $P_f(\Delta)$ can only be reached by removing these ``hopping'' segments of the particle trajectories (see SI Sec.\;IIA for details). Not only does $\tilde{\varphi}_{\mathrm{d}}$ then appears at higher densities, but as long as the network of connected cages percolates dynamical arrest is formally impossible. In that context, it is interesting to note that for a prior $\tilde{P}_f(\Delta)=\delta(\Delta)$,
the first iteration of the cavity reconstruction formalism is analogous to the void (Swiss-cheese)
percolation setup for a Poisson process~\cite{EKR84}. In addition, for a non-trivial distribution of cage sizes,
thresholding volume exclusion maps cavity reconstruction onto void percolation for polydisperse spheres~\cite{vdM96} (see SI Sec.\;IIC).
This equivalence between cavity reconstruction and void percolation sheds light on the single-cage assumption.
In the iRFOT description, the MSD of each particle should remain finite when $\varphi>\varphi_{\mathrm{d}}$,
but by construction the MSD can only be truly bounded if (minimally) $\varphi>\varphi_\mathrm{p}$, the void percolation transition.
%Yet paradoxically, for all $d$ studied, we find that $\tilde{\varphi}_{\mathrm{d}}<\varphi_{\mathrm{p}}$.
%The situation is even more dire if one considers that even for $\varphi>\varphi_\mathrm{p}$,  a single-particle can hop between
%multiple cages as long as a (possibly tenuous) path connects them (Fig.~\ref{fig:hopping}a).

From MD simulations of the MK model, we
detect the first hopping event of each particle (see SI Sec.\;IIIA for details).
%to check if the process is cooperative or not.
%Indeed, the temporal distribution of hopping, by contrast to that of standard finite-dimensional HS~\cite{CDB09}, shows no hopping cascades (Fig.~\ref{fig:hopping}b).
%This analysis thus shows that $\varphi_{\mathrm{p}}>\varphi_{\mathrm{d}}$ is the only sharply-defined dynamical transition.  \added{how is this statement concluded from facilitation analysis above?}
Around $\varphi_{\mathrm{d}}$, mode-coupling and hopping processes mix, but hopping quickly dominates the dynamics upon increasing $\varphi$. Although the hopping of a particle does not leave an empty void in the MK model,  it can nonetheless unblock a channel for a neighboring particle to leave its cage and hence facilitate its hopping. Facilitation is thus present, but weaker than in standard finite-dimensional HS, especially at high densities. 
%{\color{red} 
Weakened facilitation is notably
%} 
signaled by the fact that the distribution of hopping times
computed 
%{\color{red} 
from a regular MD simulation largely
%} 
coincides with the distribution obtained in the cavity procedure, where a single particle hops in an environment
where neighboring particles are forbidden to 
%{\color{red} 
do so
%} 
(Fig.~\ref{fig:hopping}b inset). 
We find the cumulative distribution of hopping times over the accessible dynamical range to be well described by a power law $G_{\rm h}(t) = (t/\tau_\mathrm{h})^{1-\mu}$  (Fig.~\ref{fig:hopping}b),
with the characteristic hopping time $\tau_\mathrm{h}$ increasing 
%{\color{red} 
roughly exponentially with $\varphi>\varphi_\mathrm{d}$ and markedly increasing with $d$
%} 
(Fig.~\ref{fig:hopping}d). 
%{\color{red} 
This Arrhenius-like scaling form is consistent with a gradual and uncorrelated narrowing of the hopping channels with $\varphi$.
%} 
Note that similar phenomenological power-law distributions have recently been reported for other glass-forming systems, such as the bead-spring model for polymer chains~\cite{Helfferich2014}. We get back to this point in the conclusion.

\begin{figure}[t]
\includegraphics[width =\columnwidth]{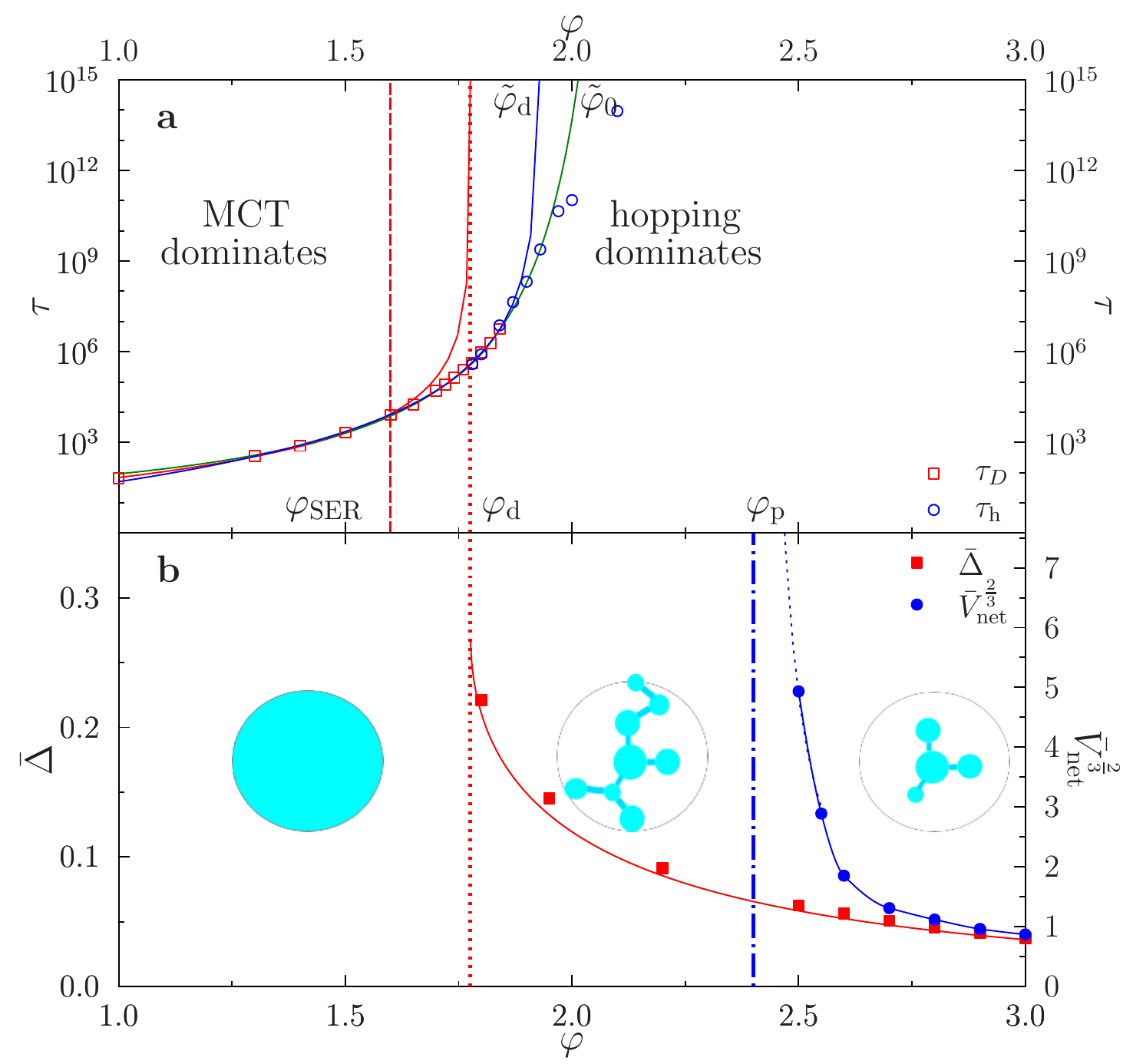}
\caption{(a) Dynamical and (b) static phase diagrams for the MK model in $d=3$. 
%{\color{red} 
Early in the critical regime, the relaxation times scale like a power-law,
%}
 but beyond $\varphi_{\mathrm{SER}}$ hopping causes large deviations from this scaling. An effective $\tilde{\varphi}_{\mathrm{d}}$ is numerically detected instead. A VTF scaling fits the data even better. Statically, cages can be detected from $\varphi_{\mathrm{d}}$ onwards by removing hopping. In reality, the fine inter-cage channels that allow hopping, however, result in a cage network. Beyond $\varphi_{\mathrm{p}}$ the typical network stops percolating and 
%{\color{red} 
the network volume  scales critically $\bar{V}_{\mathrm{net}}\sim (\varphi-\varphi_p)^{-1.8}$ with $\varphi_{\mathrm p}=2.40$ (dashed blue line)~\cite{EKR84,SA94}.
%}
The single-cage limit is reached when $\bar{V}_{\rm net} \sim \bar{\Delta}^{3/2}$. \label{fig:PD}}
\end{figure}

\paragraph*{Finite-dimensional phase diagram - }
A clear scenario for hopping in the MK model follows~(Fig.~\ref{fig:PD}).
Dynamically, the system becomes increasingly sluggish upon increasing $\varphi$ above $\tilde{\f}_{\rm onset}$.
Initially, cages are not well formed and the slowdown exhibits a power-law scaling, according to the iRFOT critical predictions.
Hopping cannot be defined because cages are too loose.
Upon approaching $\f_{\rm d}$, however, cages become much longer-lived. In this regime, iRFOT predictions give a rapidly growing $\t_D$, but
hopping processes allow particles to escape their cages and diffuse, hence providing a cutoff to the critical divergence of $\t_D$.
The critical-like behavior of the diffusivity is also pushed to denser systems, and fitting to a power-law gives
 $\tilde{\varphi}_{\mathrm{d}} >  \f_{\rm d}$.
When $\tau_{D}$ is comparable to $\t_{\rm h}$
a mixed regime emerges, characterized by a SER breakdown, as we discuss below.
Even beyond $\tilde{\varphi}_{\mathrm{d}}$, however, the dynamics is not fully arrested. Hopping remains possible, which shows that $\tilde{\varphi}_{\mathrm{d}}$
has no fundamental meaning and is just a fitting parameter associated to an effective power-law divergence of $\t_D$.
In fact, the MK dynamical data are better fitted by a VTF form than by the critical power-law (Fig.~\ref{fig:PD}a),
%{\color{red}
although
 the fitting parameter $\varphi_{0}$ has no direct static interpretation because it is intermediate between
$\varphi_{\mathrm{d}}$ and $\varphi_{\mathrm{p}}$.
%}

The dynamics can also be understood from the organization of cages.
The critical density $\varphi_{\mathrm{d}}$ of iRFOT corresponds to the emergence of a connected network of cages.
Typical networks for $\varphi_{\mathrm{d}}<\varphi<\varphi_{\mathrm{p}}$ span the system volume. When $\varphi>\varphi_{\mathrm{p}}$, they become finite and the mean network volume $\bar{V}_{\rm net}$ (sum of cage volumes in the network) follows a critical scaling from standard percolation (Fig.~\ref{fig:PD}b).
Based on this analysis, 
%{\color{red} 
in the absence of facilitation the dynamical arrest
%} 
should take place at $\varphi_{\mathrm{p}}$~\cite{HFF06}.
Note that although above $\varphi_{\mathrm{p}}$ the single-particle MSD is bounded,
a particle can still explore a finite number of cages. Perfect single-cage trapping can only be found at $\varphi\to\io$ in finite $d$. Hopping is then infinitely suppressed because  both the width and the number of hopping channels between cages vanish.
However, even if hopping interferes with caging, well above $\varphi_{\mathrm{d}}$ vibrational relaxation within the cage is sufficiently quick to numerically distinguish it from hopping. This large separation of timescales enables the facile detection of hopping in MD simulations and cavity reconstruction. But upon approaching $\varphi_{\mathrm{d}}$ the task becomes acutely sensitive to the arbitrary thresholding inherent to any hopping detection algorithm~\cite{VKBZ02,CDB09} (see SI Sec.\;IIIA for details).

%From the liquid dynamics point of view, hopping results in a ``renormalization'' of the dynamical transition $\tilde{\varphi}_{\mathrm{d}}$ and power-law exponent $\tilde{\gamma}$ with respect to their mean-field estimates, $\varphi_{\mathrm{d}}$ and $\gamma$, respectively.

As expected from the exactness of the iRFOT description in $d\rightarrow\infty$, $\tilde{\varphi}_{\mathrm{d}}/\varphi_{\mathrm{d}}\rightarrow 1$ with increasing $d$. Both $\tilde{\gamma}$ and $\gamma$ also appear to converge to the $d=\infty$ value (Fig.~\ref{fig:MSD}b)~\cite{KPUZ13}. Because $\varphi_{\mathrm{d}}<\varphi_{\mathrm{p}}$ for all $d$, the suppression of hopping with increasing $d$ 
%{\color{red} 
(see Fig.~\ref{fig:MSD}d inset)
%} 
ought to be ascribed either to the narrowing of the hopping channels or to topological changes to the cage network. Because the pressure at the dynamical transition increases only slowly with dimension ($p_\mathrm{d}\sim d$), the typical channel width is expected to stay roughly constant. The topology of the cage network, however, has a larger dimensional dependence. The cage network at percolation, for instance, has a fractal dimension $d_\mathrm{f}\ll d$~\cite{SA94}, e.g, $d_\mathrm{f}=4$ for $d \ge d_u = 6$. Although this result is only valid at $\varphi_\mathrm{p}$ proper, the \emph{local} network structure persists at smaller $\varphi$ because the loss of the cage network fractality takes place through the single-point inclusion of non-percolating clusters~\cite{SA94}. The network topology is therefore such that the hopping channels (even assuming that their cross-section remains constant) cover a vanishingly small fraction of the cage surface as $d$ increases. The limited number of ways out of a local cage thus entropically suppresses hopping.

\begin{figure}[t]
\includegraphics[width =\columnwidth]{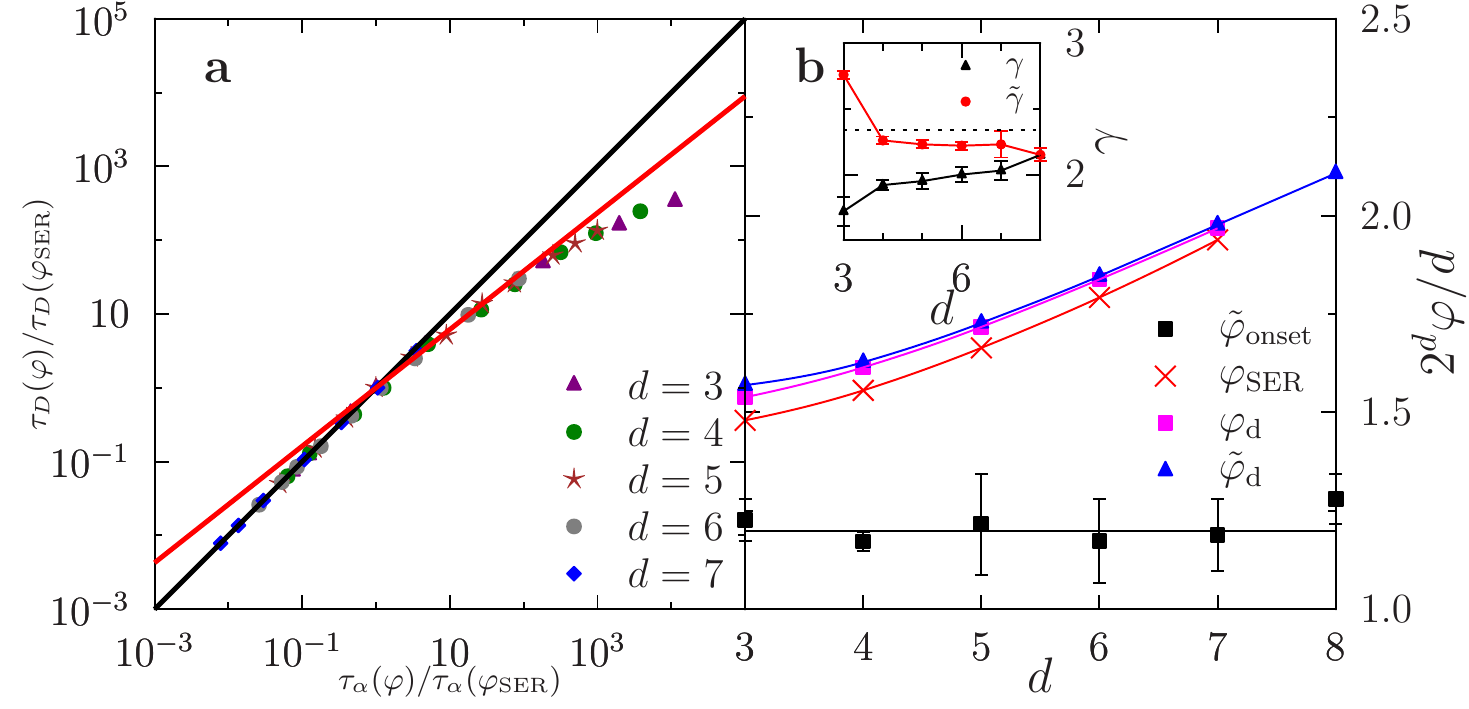}
\caption{(a) Dimensional rescaling of the SER (black) and SER breakdown regimes for standard finite-dimensional HS. The early deviation exponent $\omega$ is consistent with hopping in the MK model with $\omega=0.22$ (red line, see Fig.~\ref{fig:MSD}d), but a growing deviation is observed as $\varphi$ increases. (b) The dimensional scaling of HS results for $\tilde{\varphi}_\mathrm{d}$, $\varphi_\mathrm{d}$, and $\varphi_{\mathrm{SER}}$ converge as $d$ increases, while $\tilde{\varphi}_{\mathrm{onset}}$ remains distinctly smaller (compare with Fig.~\ref{fig:MSD}c). Note that
in $d=8$,  $\tilde{\varphi}_\mathrm{d}$, $\varphi_\mathrm{d}$, and $\varphi_{\mathrm{SER}}$ are numerically indistinguishable. (inset) Dimensional evolution of $\gamma$ and $\tilde{\gamma}$. Both of which are consistent with the $d=\infty$ result (dashed line). Solid lines are guides for the eye. \label{fig:HS}}
\end{figure}

\paragraph*{SER Breakdown -} With hopping events clearly identified, it becomes possible to isolate the pure critical iRFOT (or mode-coupling) regime. 
Within this regime, we obtain a power-law scaling that is consistent with $\varphi_\mathrm{d}$ (see SI Sec.\;IIB for details), and the SER is followed. Deviations from the extrapolated critical scaling coincide with the SER breakdown in all $d$.  Although $\tilde{\varphi}_{\mathrm{onset}}$ occurs at a roughly constant distance from $\varphi_{\mathrm{d}}$, the SER breakdown occurs in systems that are increasingly sluggish with $d$, $\varphi_{\rm SER} \to \varphi_{\rm d}$, and thus properly converges to the idealized mean-field behavior as $d\rightarrow\infty$.  In the MK model, the SER breakdown is thus clearly due to hopping.

By modifying the cavity reconstruction analysis, a self-consistent caging determination of $\f_\mathrm{d}$ and $\f_{\rm p}$ should also be possible for standard finite-dimensional HS. We do not attempt such a computation here, but instead use the insights gained from the MK model to associate the SER breakdown in HS with hopping. We fit the dynamical data from the regime
over which the SER is obeyed to extract $\varphi_{\mathrm{d}}$ and $\gamma$, and the full dynamical regime to extract $\tilde{\varphi}_{\mathrm{d}}$ and $\tilde{\gamma}$~\cite{CIPZ11}.
As for the MK model, the two procedures converge as $d$ increases (Fig.~\ref{fig:HS}), while $\tilde{\varphi}_{\mathrm{onset}}$ clearly remains distant, as is observed in many other glass formers~\cite{FSS13,HBKR14}.
Interestingly, for HS, $\varphi_{\mathrm{SER}}$ and $\varphi_{\mathrm{d}}$ are relatively close to begin with.
The fairly structured pair correlation function in HS and the much larger pressure at $\varphi_{\mathrm{d}}$
lead to smaller interparticle gaps. Particles are thus caged more efficiently, which suppresses hopping.

Contrasting Figs.~\ref{fig:MSD}d and~\ref{fig:HS}a suggests that near $\varphi_{\mathrm{SER}}$ the SER breakdown exponent $\omega$ is similar for HS and the MK model.
In this regime, HS hopping is consistent with MK-like hopping. In HS, however, single-particle hopping leaves an actual structural void that enhances the correlation
(and hence the facilitation) of hopping events~\cite{GC03,CWKDBHR10,KHGGC11}.
As HS become more sluggish, cooperativity plays a growing role.  As a result, a pronounced difference between HS and MK hopping for
$\varphi\gg\varphi_{\mathrm{SER}}$ can be observed.
The lack of a notable dimensional dependence of the master curve suggests that if the SER breakdown is also affected by critical fluctuations,
as suggested in  Ref.~\cite{BB07}, that effect may be hard to detect. In contrast to Ref.~\cite{CCJPZ13}, we now understand the reduction of the
measured $\omega$ as $d$ increases to a delayed onset of hopping.

\paragraph*{Conclusions - }
We have numerically and theoretically studied a model glass former 
in which it is possible to isolate hopping from the critical mode-coupling dynamical slowing down, and in which no other dynamical effects are present besides these two.
The results illuminate the key role played by hopping in suppressing the iRFOT dynamical transition in finite-$d$ and in breaking the SER scaling.  
%{\color{red} 
The MK model gives an example where single-particle hopping is sufficient to cause the SER breakdown, but in HS facilitation likely amplifies the effect, 
which could explain the dependence of $\omega$ on density (Fig.~\ref{fig:HS})~\cite{HMCG07}.
%}
%The analysis further suggests that an \emph{ab initio} theory of hopping in the MK model should be within reach. 
%Analytic computations might also be performed from dynamical generalizations of the cavity method.

%{\color{red}
 For standard finite-dimensional HS and other structural glass formers, we expect the situation to be made more complex by the other dynamical processes mentioned in the introduction.
One might then conjecture the existence of at least three dynamical regimes for glass formers, upon increasing density.
(i)~A iRFOT/mode-coupling regime below $\varphi_{\mathrm{SER}}$.
(ii)~A MK-like hopping regime around $\f_{\rm SER}$, where hopping is the dominant correction to the iRFOT description,
the mode-coupling critical scaling holds but the apparent mode-coupling
transition shifts to higher densities and the effective exponent $\g$ changes,
and the SER breakdown is incipient.
In this regime the hopping timescale increases (exponentially) quickly with density (Fig.~\ref{fig:hopping}d).
We expect this increase to be similar for HS and MK liquids, because the probability of finding a neighboring
cage is roughly $\exp(-\f)$ for both models. 
(iii)~At yet higher densities, hopping becomes too slow and
other dynamical effects likely become important.
If glass-glass nucleation barriers do not grow as quickly as the hopping barriers, 
then these processes may eventually become the dominant relaxation mechanism,
following the RFOT prediction~\cite{KTW89,XW01a,BB04}. In this regime (hence in deeply supercooled liquids much below $T_{\rm d}$) 
the VTF law and the associated Adam-Gibbs relation should be  reasonably well obeyed.
Note that other processes such as cooperative hopping dressed by elasticity might also occur in this regime~\cite{MS13}.
Note also that these different regimes are probably not separated by sharp boundaries in realistic systems, 
hence all these relaxation processes might coexist, making their identification quite challenging.
%}

We would also like to stress, in line with previous studies, that VTF fits of the structural relaxation time
in regimes (i) and (ii) should not be used
to extract the putative Kauzmann transition point. In our opinion it makes no sense to test the Adam-Gibbs relation
in these dynamical regimes. In the MK model, although the VTF law can be used to fit the dynamical data,
 there is indeed no associated Adam-Gibbs relation and  $\varphi_{0}$ has no thermodynamic meaning. In particular,
$\varphi_0$ is {\it not} associated with a Kauzmann transition (which in the MK model only happens at $\f=\io$~\cite{MK11}).
This observation is particularly important for numerical simulations and experiments on colloids and granular systems,
which are most often performed in the vicinity of $\f_{\rm d}$ and $\f_{\rm SER}$, and hence are found within the first two regimes.

%{\color{red} 
Finally, we note that the MK model could also serve as a test bench for descriptions of hopping~\cite{Sc05,CBK07,MMR06}, as well as for relating percolation and glassy physics more broadly~\cite{ACFS14}. These studies may clarify other finite-dimensional effects, such as the correlation observed between local structure and dynamics~\cite{TKSW10}.
{\bf Acknowledgments}
We thank G.~Biroli, J.-P.~Bouchaud, D.~Chandler, J.-P.~Garrahan,
J.~Kurchan, D.~Reichman, C.~Rycroft, and G.~Tarjus for stimulating discussions. Financial support was pro- vided by the European Research Council through ERC grant agreement no. 247328 and ERC grant NPRGGLASS. P.C. acknowledges support from the Alfred P. Sloan Foundation. %and from National Science Foundation Grant no. NSF DMR-1055586.
%\end{acknowledgments}
\clearpage
%% PNAS does not support submission of supporting .tex files such as BibTeX.
%% Instead all references must be included in the article .tex document.
%% If you currently use BibTeX, your bibliography is formed because the
%% command \verb+\bibliography{}+ brings the <filename>.bbl file into your
%% .tex document. To conform to PNAS requirements, copy the reference listings
%% from your .bbl file and add them to the article .tex file, using the
%% bibliography environment described above.

%%  Contact pnas@nas.edu if you need assistance with your
%%  bibliography.

% Sample bibliography item in PNAS format:
%% \bibitem{in-text reference} comma-separated author names up to 5,
%% for more than 5 authors use first author last name et al. (year published)
%% article title  {\it Journal Name} volume #: start page-end page.
%% ie,
% \bibitem{Neuhaus} Neuhaus J-M, Sitcher L, Meins F, Jr, Boller T (1991)
% A short C-terminal sequence is necessary and sufficient for the
% targeting of chitinases to the plant vacuole.
% {\it Proc Natl Acad Sci USA} 88:10362-10366.

%% Enter the largest bibliography number in the facing curly brackets
%% following \begin{thebibliography}

\begin{widetext}

%\bibliographystyle{mioaps}
%\bibliography{HS,glass,hopping,hopping_SI}
%\bibliography{HS,glass}

\appendix

\centerline{\large \bf Supplementary Information}
\tableofcontents 

\section{Introduction}
\subsection{The model}

The 
%{\color{red} 
infinite-range variant of the Mari-Kurchan (MK) model~\cite{MK11} is defined by adding to the distances between pairs of particles an additional quenched random shift that spans the whole system size (Fig.~\ref{fig:model}a). The Hamiltonian for $N$ hard spheres (HS)
%} 
is thus \beq \mathcal{H} =
\sum_{i<j}^{N} U(|\br_i - \br_j + \bm{\L}_{ij}|), \eeq where $U(r)$ for $|\br| = r$  is the HS potential ($e^{-\b U(r)} = \theta(r-\sigma)$), for spheres of diameter $\sigma$, and $\bm{\L}_{ij}$
is a uniformly distributed vector within the system volume $V$, i.e., with a probability distribution $P(\bm{\L}_{ij}) = 1/V$.
Note that the standard HS model corresponds to $\bm{\L}_{ij}=0$. 

%{\color{red} 
Note that even if in principle all particles interact with all others, in practice because $U(r)$ is short-ranged, a given particle only interacts directly with a finite number of neighbors (the first coordination shell), as in usual liquids. Hence, the model is akin to a mean-field spin-glass model with finite connectivity with the connectivity depending on the number of neighbors in the first coordination shell, and thus on both density and dimension.
%}

MK liquids have a simple structure in all spatial dimension $d$, because random shifts eliminate higher-order correlations. For example, consider two particles $j$ and $k$ both near particle $i$, i.e., $|\br_i
- \br_j + \bm{\L}_{ij}| \approx \sigma$ and $|\br_i - \br_k + \bm{\L}_{ik}| \approx \sigma$. Unlike in regular HS, in the MK model particles $j$ and $k$ have a negligible probability of being
near each other ($|\br_j - \br_k + \bm{\L}_{jk}| > \sigma$), because their effective distance is shifted by $\bm{\L}_{jk}$, which is of the order of the system size. This argument can also be
generalized to interactions between more particles. Particle $i$ thus has hard-core interactions with its neighbors, but with probability one in the thermodynamic limit these neighbors can overlap
with each other. 
%The structure of the system is encoded by the $n$-point structural correlations, that should however be defined by properly taking into account
%the random shifts.
The two-point correlation function seen from one particle is simply
\begin{equation}
g_2( \br) = 
\ol{\left\langle \frac1N \sum_{i \neq j} \d( \br_{ij}- \br ) \right\rangle^{\bm\Lambda}} = 
\theta (|\br| - \sigma) \ , 
\label{eq:g2}
\end{equation}
where  $\br_{ij}=\br_i - \br_j + \bm{\L}_{ij}$.  

%while the $n$-point function is defined as
%\begin{equation}
%g_n( \br_1 \cdots \br_{n-1} ) = \left\langle \frac1{N^{n-1}} \sum_{i_1 \neq i_2 \neq \cdots i_n} 
%\d( \br_{i_1i_2} - \br_1 ) 
%\d( \br_{i_2i_3}- \br_2 ) 
%\cdots
%\d( \br_{i_{n-1}i_n} - \br_{n-1} ) 
%\right\rangle  \ .
%\end{equation}
%Because of the discussion above one has
%\begin{equation}\label{eq:gn}
%g_n(\br_1, \br_2, \ldots, \br_n) = \prod_{ j=1\ldots n;j\neq i} g_2(\br_i,\br_j), \quad \forall n>2,
%g_n(\br_1, \br_2, \ldots, \br_n) = \prod_{ 1\leq i<j\leq n} g_2(\br_{ij}), \quad \forall n>2.
%\end{equation}
%for any $i$, where $r_{ij} = |\br_i - \br_j|$.
%Note that the MK model therefore has the same liquid structure as HS in $d\rightarrow \infty$~\cite{FP99,PS00,PZ10,TS10,MK11}.
%{\bf (Here I want to say MK is equivalent to the HS in large d, but I am not sure. I think Eq. (3) is true, but I don't know if Eq. (2) is true around $\ph_d$).}

Let $V_{\rm d}(\sigma/2)$ be the volume of a $d$-dimensional ball of diameter $\sigma$, and $V_{\rm d} = V_{\rm d}(1)$.
For the MK model, the virial expansion of the equation of state (EOS) terminates at the second-order \beq
\begin{split}
p &= 1 + B_2 \rho = 1 + 2^{d-1} \ph \ , \\
S^{\rm MK}_{\rm liq} & = 1 - \log\r\lambda^d  - 2^{d-1} \f +\ln N, \label{eq:liquidEOS}
\end{split}
\eeq where $p = \beta P/ \rho$  is the reduced pressure with $\beta$ the inverse temperature and $\rho=N/V$ the number density, $\ph=\rho V_d(\sigma/2) = \rho V_d 2^{-d}$ is the packing fraction (we set $\sigma = 1$), $S^{\rm MK}_{\rm liq}$ is the liquid
entropy per particle, $\lambda$ is the thermal de Broglie wavelength, and $B_2 = V_d /2$ is the HS second virial coefficient.

\begin{figure}[!h]
\centerline{ \includegraphics [width = 2.2in] {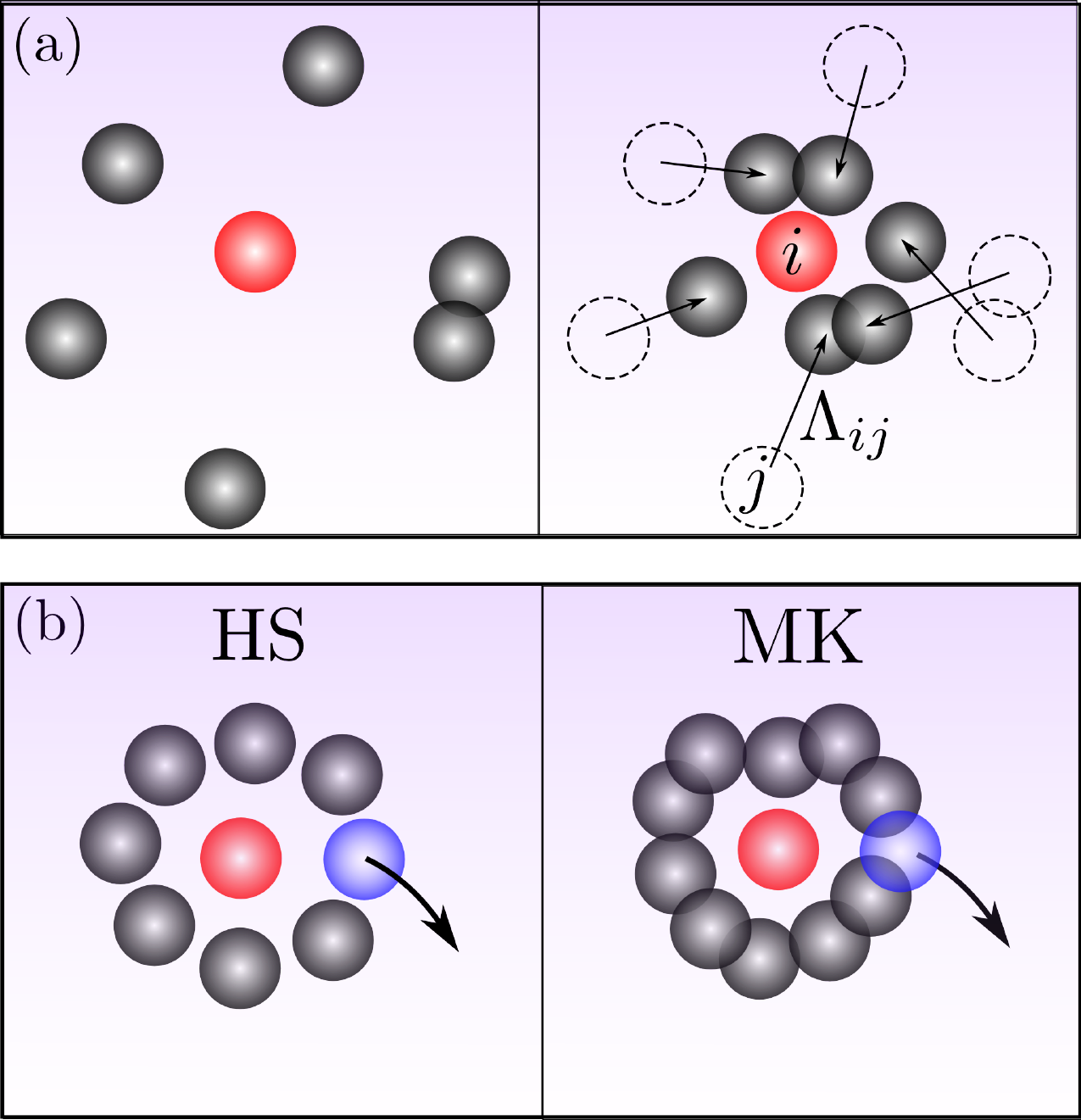}} \caption{(a) Illustration of the MK model. Left: particles in the original space. Right: particles in the shifted space with respect to
particle $i$ (red). Although neighbors cannot overlap with particle $i$, they are allowed to overlap with each other because of the random shifts. (b) Comparing hopping in the HS and the MK models. Left: in HS, removal (hopping) of a neighbor (blue)
creates an open channel for the caged particle (red) to hop.  Right: In MK, removing a neighbor is much less likely to open a channel because the other neighbors are allowed to overlap.} \label{fig:model}
\end{figure}

Compared to HS, the MK model has several unique features. (i) Although monodisperse HS easily crystallize in low dimensions, the random shifts in the MK model impose a quenched
disorder that is incompatible with crystal symmetry and fully suppresses the crystal phase. (ii) Glass-glass nucleation~\cite{XW01a,BB04} is also suppressed, because if a nucleus forms
around a particle, the particles inside this nucleus are actually randomly distributed in real space and thus no surface can be formed (Fig.~\ref{fig:model}). The free energy cost of
forming a nucleus hence scales with the system size and diverges in the thermodynamic limit. (iii) Particle hopping is much less correlated. By contrast to HS, where the hopping of a particle
increases the chance that one of its neighbors also hops due to the real-space void it leaves behind, facilitation is limited to unblocking an escape channel in the MK model (Fig.~\ref{fig:model}b). (iv) MK
particles are distinguishable because the quenched shifts $\{\bm{\L}_{ij}\}$ are fixed. Besides the lack of structure, the partition function $Z$ of the MK model is therefore different
from that of HS by a factor of $N!$, i.e., $Z_{\rm MK}/N! \sim Z_{\rm HS}$, and hence $S^{\rm MK}_{\rm liq} \sim S_{\rm liq}^{\rm HS} + \ln N$. As a result, the density of the Kauzmann
transition in MK diverges in the thermodynamic limit~\cite{MK11}.

Introducing a set of quenched random shifts brings two key advantages from a methodological point of view. First, in computer simulations, it is convenient to ``plant" an
equilibrated MK configuration (Sec.~\ref{sec:planting}). Planting avoids the (circular) difficulty encountered in most other glass-forming liquids of equilibrating an initial liquid configuration before studying its equilibrium relaxation dynamics. %In MK, one can plant liquid configurations at arbitrary densities, close or well above the dynamical transition, a feat that would otherwise be nearly impossible by simple annealing.
Second, one can map the model onto a constraint satisfaction problem defined on a random graph (Bethe lattice). It is therefore possible to study its properties with the cavity
method (Sec.~\ref{sec:cavity}), which is, in principle, exactly solvable.

%\begin{figure}
%\centerline{ \includegraphics [width = 2.1in] {hopping_illustration}} \caption{Comparing hopping in HS and the MK models. (a) In HS, removal (hopping) of a neighbor (blue)
%creates an open channel for the caged particle (red) to hop. (b) In the MK model, removing a neighbor is much less likely to open a channel because the other neighbors are allowed to overlap.}
%\label{fig:hopping_illustration}
%\end{figure}

\subsection{Simulation details}

\subsubsection{Molecular dynamics simulations}
We adapt the event-driven molecular dynamics (MD) algorithm of Refs.~\cite{SDST06, CIPZ11, CIPZ12} for HS to simulate the MK model in dimensions $d= 2-6$ with $N=4000$ particles. %Here the particle-particle interactions should be calculated on the shifted displacement $|\br_i - \br_j + \Lambda_{ij}|$.
Periodic boundary conditions with the minimum image convention are implemented on the shifted distances $|\br_i - \br_j + \bm{\Lambda}_{ij}|$. For each $\varphi$, we perform 8
independent realizations, each corresponding to a different set $\{\bm{\Lambda}_{ij}\}$ for a planted initial configuration (see Sec.~\ref{sec:planting}).
Simulations are run at constant unit $\beta$, for a time $t$ (given in units of $\sqrt{\beta m \sigma^2}$, where the particle mass $m$ is also set to unity) sufficiently long to reach either the diffusive regime in the liquid or the asymptotic plateau in the glass. %As in the case of HS, the finite-size effects are negligible in large dimensions even like $d=6$, because the correlation length shrinks rapidly with $d$.
As described in Refs.~\cite{CIPZ12,CCJPZ13}, HS data is obtained from simulations of $N=8000$ identical particles in $d=4-8$, and, in order to prevent the system from crystallizing~\cite{ZSWLSSO14}, from a HS binary mixture with diameter ratio $\sigma_2/\sigma_1 = 1.2$ in $d=3$~\cite{CCJPZ13} . %monodisperse systems of hard spheres are usually partially crystallized. To avoid this problem, in $d=3$ we . For each density, 8 independent configurations are simulated.

\subsubsection{Planting}
\label{sec:planting}

Planting, which here consists of switching the order of determining initial particle positions $\{\br_i\}$ and constraints $\{\bm{\L}_{ij}\}$, is an expedient
technique for studying equilibrium ensembles in random constraint satisfaction problems~\cite{KZ09}.
In general, the planted ensemble is different from the annealed ensemble, but for the liquid phase it can be shown that both are equivalent, as we detail below.

In the following, we will be interested in physical observables $\mathcal{F}$ that depend
on some initial condition $\{\br_i\}$ and on their time evolution under deterministic
MD dynamics, e.g. the mean square displacement defined in Eq.~\eqref{eq:MSDdef}.
In the presence of disorder, the average of physical observables should be measured by the so-called ``quenched'' average
\begin{equation}
\ol{\la \mathcal{F} \ra ^{\bm{\L}}} \equiv \int \prod_{i<j}^{N} d\bm{\L}_{ij} P(\bm{\L}_{ij}) \left( \frac{\int \prod_{i=1}^{N} d\br_i \mathcal{F} e^{-\beta\mathcal{H}}}{\int
\prod_{i=1}^{N} d\br_i  e^{-\beta\mathcal{H}}}\right), \label{eq:quenchF}
\end{equation}
where $\mathcal{F}$ and $\mathcal{H}$ depend on both $\{ \br_i \}$ and $\{ \bm{\L}_{ij} \}$.
In fact, because the disorder is independent of time for a given sample, one should first perform the thermal ensemble average
${\la \mathcal{F} \ra}$ for a given realization of disorder, and then repeat this operation
for many extractions of $\{\bm{\L}_{ij} \}$
 to average over the disorder.  In simulations, however, once $\{ \bm{\L}_{ij} \}$ is fixed, equilibrating independent configurations at large $\ph$ is very time consuming, because one should first anneal the system quasi-statically slowly up to the desired density. %, and then wait sufficiently long time until it reaches equilibrium.

Let us define the so-called ``annealed'' average:
\beq  \label{eq:annealF}
\la \mathcal{F} \ra_{\rm a} \equiv \frac{\int \prod_{i<j}^{N} d\bm{\L}_{ij} P(\bm{\L}_{ij})  \int
\prod_{i=1}^{N} d\br_i \mathcal{F} e^{-\beta\mathcal{H}}}{\int \prod_{i<j}^{N} d\bm{\L}_{ij} P(\bm{\L}_{ij}) \int \prod_{i=1}^{N} d\br_i  e^{-\beta\mathcal{H}}} =
\frac{\int
\prod_{i=1}^{N} d\br_i \int \prod_{i<j}^{N} d\bm{\L}_{ij} P(\bm{\L}_{ij}) e^{-\beta\mathcal{H}} \mathcal{F} }
{\int
\prod_{i=1}^{N} d\br_i \int \prod_{i<j}^{N} d\bm{\L}_{ij} P(\bm{\L}_{ij}) e^{-\beta\mathcal{H}}}
\ .
\eeq
This average corresponds to a very different situation, where the averages over
$\{ \br_i \}$ and $\{ \bm{\L}_{ij} \}$ are interchangeable.
Physically, this describes a situation where both variables and disorder fluctuate together; their timescales are indistinguishable.
Mathematically, the last equality in Eq.~\eqref{eq:annealF} shows that the integration measure can be obtained
by first extracting a uniformly random configuration $\{ \br_i \}$, and next extracing a configuration $\{ \bm{\L}_{ij} \}$ from the distribution
\beq
P(\{ \bm{\L}_{ij} \} | \{ \br_i \}) = \left[ \prod_{i<j}^{N} P(\bm{\L}_{ij}) \right] \, e^{-\beta\mathcal{H}}
= \prod_{i<j}^{N} \left[  P(\bm{\L}_{ij}) e^{-\b U(|\br_i - \br_j + \bm{\L}_{ij}|)} \right] \ .
\eeq
Because $P(\{ \bm{\L}_{ij} \} | \{ \br_i \})$ is factorized, each $\bm{\L}_{ij}$ must be extracted independently, uniformly in the volume $V$
with the constraint that $|\br_i - \br_j + \bm{\L}_{ij}| > \s$.
In summary, we use
the following procedure to compute $\la \mathcal{F} \ra_{\rm a}$:

\vskip10pt \noindent \noindent\rule{8cm}{0.4pt}
%------------------------------------------------------------------------------------------------------------------------------------------------------
\\
{\bf Procedure-Planting-MK}

\begin{enumerate}
\item  Generate $N$ particle positions  $\{ \br_i \}$ according to a Poisson (ideal gas) process.

\item For each pair of particles $i$ and $j$, randomly and independently draw a vector $\bm{\L}_{ij}$, uniformly in the sub-region of the whole volume $V$ that is
compatible with $\br_i$ and $\br_j$, $|\br_i - \br_j + \L_{ij}| > \sigma$.

\item Starting from the state given by $\{ \br_i \}$, and for the given $\{ \bm{\L}_{ij} \}$, compute the time evolution $\{ \br_i(t) \}$
from MD simulations. From this trajectory compute $\mathcal{F}$.

\item Repeat (1-3) to average over disorder and initial configurations.
\end{enumerate}
\noindent\rule{8cm}{0.4pt}
%------------------------------------------------------------------------------------------------------------------------------------------------------
\\

%If steps (1) and (2) are inverted, standard averaging is recovered. For cases where step (2) is straightforward, such as for the MK model, measuring the annealed average by
%planting is much more computationally efficient than standard annealing, but if it is hard, such as for HS, standard annealing is unavoidable. {\color{red} FZ, check that annealed/quenched discussion is not overly confusing}
%

The key to the success of this approach is determining if, and under what conditions,
the quenched and the annealed averages over the disorder are the same, $\ol{\la \mathcal{F} \ra ^{\bm{\L}}} = \la \mathcal{F}
\ra_{\rm a}$. Equations~\eqref{eq:quenchF} and~\eqref{eq:annealF} coincide if
the equality $\ol{\log Z^\L} = \log \ol{Z^\L}$ holds, where $Z= \int \prod_{i=1}^{N} d\br_i  e^{-\beta\mathcal{H}}$ is the
partition function for given $\{ \bm{\L}_{ij} \}$~\cite{MPV87,CC05,KZ09}.
This situation arises if the fluctuations of
$Z$ induced by the fluctuations of quenched disorder $\{ \bm{\L}_{ij} \}$ are very weak in the thermodynamic limit.
This condition is satisfied in the liquid phase, but is violated in the glass phase away from the equilibrium liquid line~\cite{MPV87,CC05,KZ09}.

According to the analysis of Ref.~\cite{KZ09},
in order to check numerically that the annealed and the quenched average coincide, one should compute the vibrational
(internal) entropy of the planted glass state. This can be done for example using the procedure described in Ref.~\cite{FS02}.
If the internal entropy of the glass turns out to be larger than the liquid entropy given by Eq.~\eqref{eq:liquidEOS},
then the annealing average does not coincide with the quenched average~\cite{KZ09}. Fortunately, in the MK model
the liquid entropy per particle diverges proportionally to $\log N$ (see Eq.~\eqref{eq:liquidEOS}), while the glass entropy per particle
is finite, because particles cannot exchange (at least if one neglects hopping, as discussed below). Therefore, for $N\to\io$
the liquid entropy is always larger than the glass entropy, and the annealed average is correct. In other words,
because the Kauzmann transition for the MK model is located at infinite density~\cite{MK11}, the procedure is valid.

Numerical simulations show that the annealed average done using the planting procedure discussed above is
in perfect agreement with the liquid EOS Eq.~\eqref{eq:liquidEOS} (Fig.~\ref{fig:planting}).
This result is not a surprise, because it can easily be shown that the annealed equation leads to the same liquid EOS in Eq.~\eqref{eq:liquidEOS},
but it is a consistency test for the numerical procedure. Note that the pressure remains stable over time, as it should be if one initializes
the MD simulation in an equilibrium configuration.

\begin{figure}[ht]
\centerline{\includegraphics [width = 2.5in] {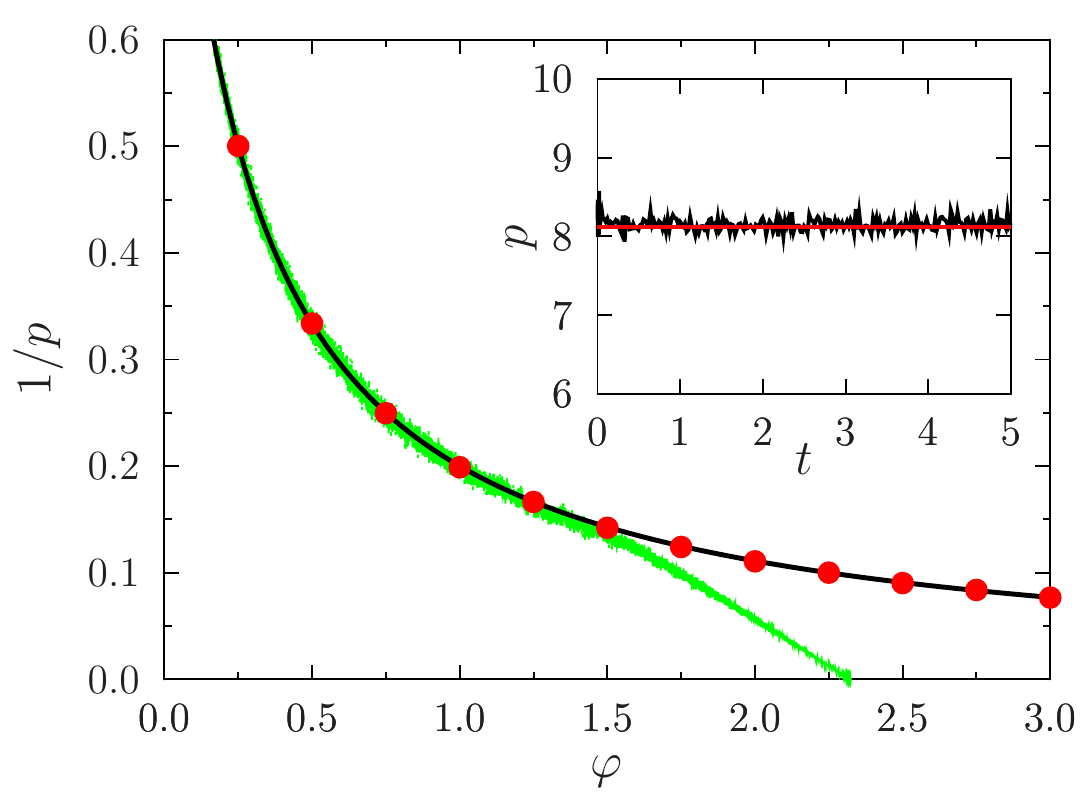}} \caption{
Comparison between the reduced pressure $p$ of planted states (red circles) and the liquid EOS Eq.~\eqref{eq:liquidEOS} (black solid line) in $d=3$. The
regime above $\varphi_{\rm d} =  1.776$ is numerically inaccessible from conventional slow quenching procedures (see, for example, the green line obtained from the Lubachevsky-Stillinger algorithm with a growth rate $\dot{\gamma}=3\times10^{-5}$~\cite{CIPZ11}), because the system starts to deviate
from the liquid EOS around $\varphi_{\rm d}$. (inset) Planting at $\varphi = 1.78$ gives the correct equilibrium pressure (red line) from $t=0$.}
\label{fig:planting}
\end{figure}

\subsection{Basic phenomenology of glassy behavior and definitions of physical quantities}
\label{sec:phenomenology}

Before turning to a more detailed explanation of our results, in this section we summarize the main physical observables that we investigate
in this study, with a short account of their definition and of the main results.

\subsubsection{Mean square displacement (MSD) and cage sizes}
\label{sec:MSD}
Despite its trivial liquid phase, the MK model presents a complex glass-forming and glassy behavior. Above the onset $\tilde{\ph}_{\rm onset}$ of sluggish dynamics (see Sec.~\ref{sec:density} for definition), We can distinguish three main regimes in the mean square displacement:
\beq \label{eq:MSDdef}
 \D (t) = \frac{1}{N}\sum_{i=1}^{N} \la |\br_i(t) - \br_i(0)|^2 \ra
\eeq
(i) a ballistic regime with $\D(t)  = d
t^2$; (ii) a caging regime with a plateau $\D
(t) \sim \bar{\D}$, where $\bar{\D}$ is the mean cage size; and (iii) a diffusive regime with $\D(t) = 2d Dt$, where $D$ is the diffusivity. %The MSD plateau starts to develop around $\f_{\rm onset}$, and,
According to mode-coupling theory (MCT), the plateau becomes asymptotically stable beyond the dynamical transition $\varphi_{\mathrm{d}}$ (see Sec.~\ref{sec:MCT}). We can then formally define the
mean cage size as the infinite time limit of the MSD
\begin{equation}
\bar{\D} \equiv \lim_{t \to \infty} \D(t),
\end{equation}
and the individual cage size $\D_i$ of each particle $i$
\begin{equation}
\D_i \equiv  \lim_{t \to \infty} \langle |\br_i(t) - \br_i(0)|^2 \rangle. \label{eq:Ai0}
\end{equation}
From this definition and the equilibrium conditions $\la \br_i(0) \ra = \la \br_i(t) \ra$ and $\la |\br_i(0)|^2 \ra = \la |\br_i(t)|^2 \ra$, we obtain another expression for $\D_i$
\begin{equation}
\begin{split}
\D_i &=  \lim_{t \to \infty} [\langle |\br_i(t)|^2 \rangle - 2 \langle \br_i(t) \cdot \br_i(0) \rangle + \langle |\br_i(0)|^2 \rangle] \\
&= 2 \lim_{t \to \infty} [\langle |\br_i(t)|^2 \rangle - |\langle \br_i(t) \rangle|^2].\\
\end{split}
\label{eq:Ai}
\end{equation}

The definition of $\D_i$ in Eq.~\eqref{eq:Ai0} can be directly used to measure individual cage sizes in numerical simulations. Equation~\eqref{eq:Ai} also suggests that $\D_i$
is twice the variance of the distribution of particle positions within a cage.
In theoretical calculations, a cage form ansatz $f_A(\br)$ is usually used for this distribution. Two
commonly used functions are the Gaussian
\begin{equation}
f^{\mathrm{G}}_{A}(\br) = \frac{e^{-\frac{\br^2}{2A}}}{(2\pi A)^{d/2}} \ \label{eq:AGaussian}
\end{equation}
and the ball functions \beq f^{\mathrm{b}}_{A}(\br) = \frac{\theta (A - \br^2)}{V_d(\sqrt{A})}. \label{eq:Aball} \eeq Below, we use the Gaussian anzatz in the replica method
(Sec.~\ref{sec:cavity}), and both ansatz in the cavity method (Sec.~\ref{sec:cavity}). The parameter $A$ in these functions can be related to $\D_i$ using Eq.~\eqref{eq:Ai} \beq
\label{MSDGaussian} \D_i = 2dA_i \eeq for the Gaussian function and \beq \label{MSDBall} \D_i= \frac{2d}{d+2} A_i
 \eeq
for the ball function.

\subsubsection{Characteristic timescales}

%{\color{red} 
In this subsection, we define the characteristic timescales, their physical interpretations, and how they are numerically 
determined.
%}

\begin{itemize}

\item 
%{\color{red} 
$\tau_0$ -- microscopic time.  This natural timescale serves as reference to compare the evolution of other timescales with spatial dimension $d$. Its definition is such that the velocity autocorrelation function $d(\tau_0)=1/e$, where $d(t) = \frac{1}{dN}\sum_{i=1}^{N}\langle  \mathbf{v}_i(t) \cdot \mathbf{v}_i(0) \rangle = \frac{1}{2d} \frac{d^2 \Delta(t)}{d t^2}$ (see Fig.~\ref{fig:tau}c)~\cite{Ikeda2013}.
%}

\item 
%{\color{red} 
$\tau_D$ -- diffusion time. The characteristic time for diffusion time is defined as $\tau_D = \sigma^2/D$,
such that $\D(t)$ vs $t/\tau_D$ collapses in the caging and diffusive regimes (Fig.~\ref{fig:tau}a), as predicted by MCT (see
Eq.~\eqref{eq:below}). Using this collapse, we can determine $\tau_D$ without explicitly extracting $D$, which allows us to estimate $\tau_D$ close to the dynamical
transition, even when the fully diffusive regime itself is beyond numerical reach.
%}

\item 
%{\color{red} 
$\tau_{\rm SER}$ -- characteristic time at $\ph_{\rm SER}$, i.e., $\tau_{\rm SER} = \tau_D(\varphi_{\rm SER})$.
%}

\item 
%{\color{red} 
$\tau_\alpha$ -- structural relaxation time. In standard glass-forming liquids, $\tau_\alpha$ 
%}
is typically extracted from the decay of the self-intermediate scattering function \beq F_s(k,t) =
\frac{1}{N} \la \sum_{i=1}^{N} e^{i \bk \cdot [\br_i(t) - \br_i(0)]} \ra, \eeq such that \beq F_s(k^*, \tau_\alpha) = 1/e, \label{eq:Fs2} \eeq where $k^*$ is the first particle peak of
the structure factor \beq S(k) = 1 + \rho \int d\br e ^ {-i \bk \cdot \br} g_2(\br). \eeq
For the MK model, however, this method cannot be directly applied because the trivial structure of $g_2(r)$ (and hence of $S(k)$) leaves $k^*$ ill defined. Here, we use a slightly different,
although consistent, approach to measuring $\tau_\alpha$. We first generalize the definition of the MSD to the typical displacement of particles
\beq
r_{\rm typ}(t) = \lim_{z \to 0}
\frac{1}{N}\sum_{i=1}^{N} \la |\br_i(t) - \br_i(0)|^z \ra^{\frac{1}{z}},
\eeq
 which is the zeroth moment of the self van Hove function $G_s(r,t)$, i.e., the displacement of the
majority of particles at time $t$. In practice, to determine $r_{\rm typ}(t)$, we use $z=0.1$ which is very close to the limit $z\to 0$.
By analogy to $\tau_D$, we then determine the relaxation time $\tau_\alpha$ by ensuring that $r_{\rm typ}^2(t)$ vs $t/\tau_\alpha$ collapses in the
MSD caging regime (Fig.~\ref{fig:tau}b).
Note that $\t_\a$ is then only defined up to an overall constant that is independent of density.

For HS, this (re)definition of $\tau_\alpha$ is consistent with the traditional one, because the condition in Eq.~\eqref{eq:Fs2} is equivalent to $k r_{\rm typ}(\tau_\alpha)
\sim 1$. The length scale $1/k^*$ indeed corresponds to that of the maximum density fluctuation, which should be of the order of the typical cage diameter. The scaling $r_{\rm
typ}(\tau_\a) \sim 1/k^* \sim \sqrt{\bar{A}}$ shows that $r_{\rm typ}$ is near the caging regime at $\tau_\a$, and hence should be independent of density.
%Numerical determinations confirm that the two measurements are equivalent {\color{red} where?}.
Our estimate of $\tau_\alpha$ is therefore consistent with the proportionality relation for the viscosity
$\tau_\alpha \sim \eta$ observed in very sluggish fluids~\cite{CCJPZ13}. 
%{\color{red} 
Note that the above definitions of $\tau_\alpha$ and $\tau_D$ give additional weight to slower and faster particles, respectively. In this context, the breakdown of SER is consistent with a proportion of fast particles that is larger than expected~\cite{KSD06}.
%}

\item 
%{\color{red} 
$\tau_{\rm h}$ -- hopping time. The typical time for a caged particle to escape (see Sec.~\ref{sec:hopping} for more details).
%}
\end{itemize}
%In Table~\ref{table:time}, we summarize the definitions of the variance time scales.

\begin{figure}[ht]
\centerline{\includegraphics [width = 4.0in] {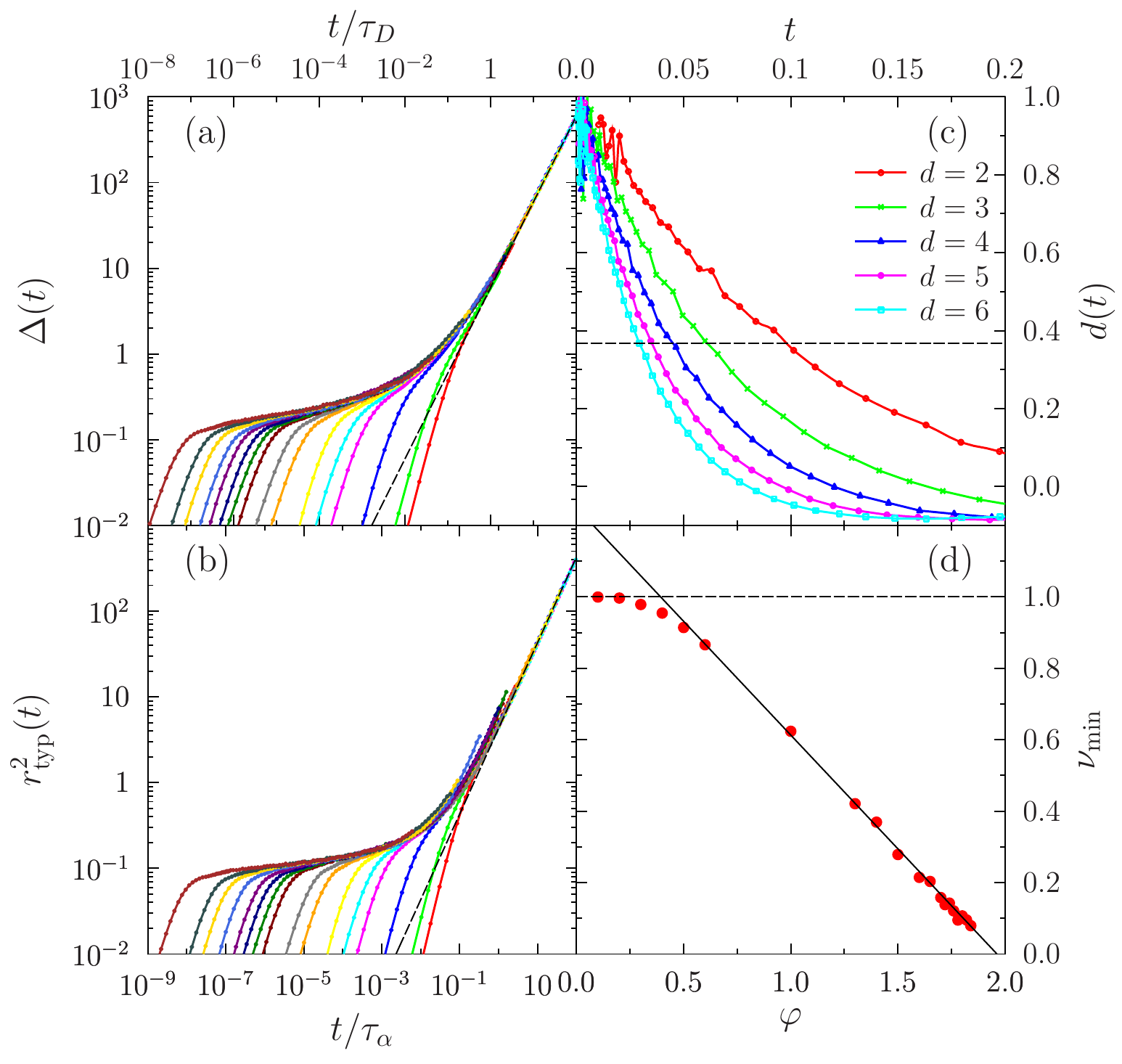} } \caption{Rescaled plots 
%{\color{red} 
of (a) the MSD and (b) the typical displacement $r_{\rm typ}^2(t)$ in $d=3$. From right to left,  $\ph=0.40,
0.60, 1.00, 1.30, 1.40, 1.50, 1.60, 1.65, 1.70, 1.72, 1.74, 1.76, 1.78, 1.80, 1.82, 1.84$, with $\t_\a$ normalized such that $\t_\a \sim \t_D$ at $\varphi_{\mathrm{SER}}$. Note that the rightmost line (red), at $\tilde{\ph}_{\rm onset}$, does not exhibit any plateau regime. (c) Velocity  autocorrelation function $d(t)$ at $\varphi_{\rm SER}$ in $d=2-6$. By definition, $d(\tau_0)=1/e$ (dashed line). (d) Minimum value of the non-Fickian coefficient $\nu_{\rm min}$ in $d=3$. At high densities the $\nu_{\rm min}$ decreases linearly with $\varphi$ (solid line), and at low densities $\nu_{\rm min}=1$ (dashed line). The crossover occurs around $\tilde \varphi_{\rm onset}  = 0.40(5)$.
%}
 }
\label{fig:tau}
\end{figure}

%\begin{table}
%\begin{ruledtabular}
%\begin{tabular}{c c}
%  notation & definition  \\
%  \hline
%  $\tau_D$ & diffusive time, $\tau_D \sim 1/D$\\
%  $\tau_\a$& relaxation time, $\tau_\a \sim \eta$\\
%  $\tau_{\rm h}$& hopping time\\
%\end{tabular}
%\end{ruledtabular}
%\caption{List of time scales used in the paper.}
%\label{table:time}
%\end{table}

\subsubsection{Characteristic densities}
\label{sec:density}
In this subsection, we define the characteristic densities (number density and volume fraction are used interchangeably), their physical interpretations, and how they are numerically and theoretically
determined. Results for HS and the MK model are reported in Table~\ref{table:phi}.
\begin{itemize}
\item $\tilde{\ph}_{\rm onset}$ -- onset density of the glassy behavior. It corresponds to
the lower limit of the caging regime~\cite{Flenner2013}.   
%{\color{red} 
Its choice is such that for $\varphi<\tilde{\ph}_{\rm onset}$ no inflection point appears in the logarithmic-scale MSD; 
for $\varphi\geq\tilde{\ph}_{\rm onset}$, the MSD shows an inflection point, i.e., a point where $\frac{d^2 \ln \Delta(t)}{(d \ln t)^2}=0$. In this
regime a non-Fickian behavior is observed. Hence,  $\tilde{\ph}_{\rm onset}$ also corresponds to
the density at which the minimum value of the non-Fickian coefficient  $\nu_{\rm min}$ is unity (Fig.~\ref{fig:tau}d), where $\nu_{\rm min} = \min_t \nu(t)$ and the 
non-Fickian coefficient $\nu(t)\equiv \frac{d \ln \Delta(t)}{ d\ln t}$~\cite{SYS08}.
%}
Note, however, that our estimate of $\tilde{\ph}_{\rm onset}$ likely underestimates the onset calculated from the emergence of a
finite configurational entropy in  static calculations~\cite{Berthier2014}. \item $\ph_{\rm SER}$ -- characteristic density for the breakdown of the Stokes-Einstein relation (SER).
Below $\ph_{\rm SER}$, hopping is irrelevant 
%{\color{red} 
because, if present, it is indistinguishable from the regular liquid dynamics, 
%} 
and the MCT scaling 
%{\color{red} 
relations
%} 
are satisfied (Sec.~\ref{sec:MCT}); above $\ph_{\rm SER}$,  hopping becomes faster than the characteristic MCT time
($\tau_{\rm h} < \tau_D$), and consequently both the MCT scaling and the SER are violated. \item $\ph_{\rm d}$ -- dynamical glass transition threshold. In the MK model, this density
is theoretically calculated from the replica method (Sec.~\ref{sec:phid}), and numerically confirmed by testing the MCT scaling $\tau_D \sim |\ph - \ph_{\rm d}|^{-\gamma}$ (or
equivalently, $D \sim |\ph - \ph_{\rm d}|^\gamma$) in the density range over which hopping is negligible ($\tilde{\ph}_{\rm onset} < \ph < \ph_{\rm SER}$). In the HS model, however, we
lack reliable theoretical predictions for $\ph_{\rm d}$ in low dimensions. We therefore determine $\ph_{\rm d}$ from fitting the simulation results for $D$ in the regime
$\tilde{\ph}_{\rm onset} < \ph < \ph_{\rm SER}$. Our results are consistent with those reported in Ref.~\cite{CIPZ11}, where $\ph_{\rm d}$ was extrapolated from slowly quenching the
fluid. Note that $\ph_{\rm d}$ is only sharply defined when hopping contributions can be separated without ambiguity. Hence, $\ph_{\rm d}$ is only well defined in the replica
calculation, where hopping is excluded by construction. \item $\tilde{\varphi}_{\mathrm{d}}$ -- effective dynamical glass transition threshold. Empirically,
$\tilde{\varphi}_{\mathrm{d}}$ is determined by fitting the diffusivity data, as is commonly done in glass formers. In this study we show, however, that $\tilde{\ph}_{\rm d}$ is
systematically shifted with respect to $\ph_{\rm d}$ ($\tilde{\ph}_{\rm d} > \ph_{\rm d}$). Note that because $\tilde{\ph}_{\rm d}$ is a fitting parameter, it also depends on the
density range one chooses (or is available) for the power-law fit.
\item $\tilde{\ph}_0$ -- phenomenological parameter from the Vogel-Tammann-Fulcher (VTF) fit to $\tau_D \sim e^{B_{\rm
VTF}/(\tilde{\ph}_0 - \ph)}$. In the MK model, $\tilde{\ph}_0$ is clearly different from the thermodynamic Kauzmann transition point $\ph_K = \infty$. Recall, however, that the MK model lacks the glass-glass
nucleation processes assumed by the Adam-Gibbs (AG) and the random first-order transition (RFOT) theories, in order to associate the divergence of the relaxation timescale with the
thermodynamic singularity at $\ph_K$. \item $\ph_{\rm p}$ -- percolation threshold for the cage network. Below $\ph_{\rm p}$, a particle can diffuse by successive hops on the
percolating network of cages. Because the infinite time limit of the MSD is only truly bounded above this threshold, $\ph_{\rm p}$ also provides a upper bound for $\ph_{\rm d}$, i.e.,
$\ph_{\rm p} > \ph_{\rm d}$. \item $\ph_{\rm K}$ -- Kauzmann transition. Density at which the complexity $\Sigma$ (or configurational entropy) vanishes. As discussed above, because
$\Sigma_{\rm MK} \sim \Sigma_{\rm HS} + \ln N$, the density of the Kauzmann transition diverges ($\ph_K = \infty$) in the thermodynamic limit.
\end{itemize}

\begin{table}
\begin{ruledtabular}
\caption{Numerical values of characteristic densities and MCT exponents for the MK and the HS models.\label{table:phi}}
\begin{tabular}{c c  c c c c c c c c c c}
 %    \multicolumn{1}{c}{}                & \multicolumn{10}{c}{MK} & \multicolumn{10}{c}{HS}\\
 & $d$ & $\tilde{\ph}_{\rm onset}$ & $\ph_{\rm SER}$ & $\ph_{\rm d}$&  $\tilde{\ph}_{\rm d}$  & $\tilde{\ph}_0$  & $\gamma$ &  $\tilde{\gamma}$ &$a$ & $b$ & $\l$ \\
  \hline
 MK &2 &0.50(5)&2.0(1) &2.398 &2.60(1)  &3.2(1)  & 4.59(4) & 5.77(4) & 0.19 & 0.26 & 0.92\\
    &3 &0.40(5)&1.60(5)&1.776 &1.93(1)  &2.15(5) & 3.27(7) & 4.95(4) & 0.25 & 0.40 & 0.85\\
  &4 &0.30(5)&1.10(5)&1.184 &1.276(2) &1.40(2) & 2.9(1) & 4.50(7)  & 0.28 & 0.46 & 0.80\\
  &5 &0.20(5)&0.70(2)&0.741 &0.783(1) &0.865(5)& 2.67(8) & 4.04(6) & 0.29 & 0.52 & 0.78\\
  &6 &0.10(5)&0.42(1)&0.445 &0.466(1) &0.510(5)&2.65(8) & 3.75(4) & 0.30 & 0.53  & 0.76\\
HS &3 &0.46(2)  &0.555(5) &0.5770(5)  &0.5885(5)  &0.603(1) &   1.72(3) & 2.8(1) & 0.40 & 1.05 & 0.47\\
   &4 &0.293(6) &0.389(6) &0.4036(2)  &0.4069(1)  &0.417(1) &   1.92(3) & 2.26(4) & 0.37  & 0.86  & 0.57\\
   &5 &0.19(2)  &0.260(5) &0.2683(1)  &0.2699(1)  &0.277(1) &   1.95(3) & 2.23(6) & 0.37 &   0.84  & 0.58 \\
   &6 &0.11(1)  &0.168(4) &0.1723(1)  &0.1731(1)  &0.178(1) &   2.00(3) & 2.22(6) & 0.36 & 0.80  & 0.60 \\
   &7 &0.065(5) &0.106(2) &0.1076(1)  &0.1081(1)  &0.112(1) &   2.0(1)  & 2.23 (7) & 0.36 & 0.80 & 0.60 \\
   &8 &0.040(2) & -       &0.06585(5) &0.06585(5) &0.0685(5)&   2.15(5) & 2.15(5) & 0.34 &  0.71 & 0.65 \\
\end{tabular}
\end{ruledtabular}
\begin{flushleft}
Data for the MK model in $d=2 - 6$ and for the HS model in $d=3-8$. Theoretical results are reported for $\ph_{\rm d}$ of the MK model, but all the other values are from simulations. In $d=8$ for the HS model, no SER violation is detected in the dynamical regime that is computationally accessible.
\end{flushleft}
\end{table}

%\begin{table}
%\begin{ruledtabular}
%\begin{tabular}{c | c c c c c}
%  $d$ & $\tilde{\ph}_{\rm onset}$ & $\ph_{\rm SER}$ & $\ph_{\rm d}$&  $\tilde{\ph}_{\rm d}$  & $\tilde{\ph}_0$ \\
%  \hline
%  3 &0.46(2)  &0.555(5) &0.5770(5)  &0.5885(5)  &0.603(1) \\
%  4 &0.293(6) &0.389(6) &0.4036(2)  &0.4069(1)  &0.417(1) \\
%  5 &0.19(2)  &0.260(5) &0.2683(1)  &0.2699(1)  &0.277(1) \\
%  6 &0.11(1)  &0.168(4) &0.1723(1)  &0.1731(1)  &0.178(1) \\
%  7 &0.065(5) &0.106(2) &0.1076(1)  &0.1081(1)  &0.112(1) \\
%  8 &0.040(2) & -       &0.06585(5) &0.06585(5) &0.0685(5) \\
%\end{tabular}
%\end{ruledtabular}
%\caption{Numerical values of the characteristic densities in HS. All the values are obtained from simulations. In $d=8$, no SER violation is detected in the dynamical regime that is computationally accessible.} \label{table:phiHS}
%\end{table}

\section{Caging}

\subsection{Thermodynamics: the caging order parameter and the dynamic transition density}

The mean caging order parameter can be obtained equivalently from the replica method, following Refs.~\cite{PZ10,BJZ11}, or from the cavity method, following
Ref.~\cite{MPTZ11}. Here we briefly describe how these approaches are adapted to the MK model.

\subsubsection{Calculation of the mean caging order parameter: the replica method}

References~\cite{PZ10,BJZ11} used the replica approach to obtain HS results, and it is straightforward to check that these derivations only rely on the pair correlation
function in the liquid phase; terms corresponding to third- and higher-order structural correlations are neglected. The treatment of Ref.~\cite{PZ10} can therefore be directly
applied to the MK model, for which these assumptions are exact.
%thanks to the factorization property of Eq.~\eqref{eq:gn}. 
The results from Ref.~\cite{BJZ11} have also been obtained using the
approximation $g_2(r) = y_{\rm liq}^{\rm HS}(\f) \th(r-\s)$, see \cite[Eq.(21)]{BJZ11} (the soft-sphere temperature is set to zero to study hard-core systems). Comparing this result with Eq.~\eqref{eq:g2}, we see that for the MK
model $y_{\rm liq}^{\rm HS}(\f)=1$. All the results of Refs.~\cite{PZ10,BJZ11} can thus be straightforwardly extended to the MK model by setting $y_{\rm liq}^{\rm HS}(\f)=1$. (Note that the
discussion of Ref.~\cite{BJZ11} was restricted to $d=3$, but a general discussion for all $d$ can be found in Ref.~\cite{PZ10}). The replicated entropy thus has the form
\beq\label{eq:SSrep}
\begin{split}
\SS(m,A;T,\ph) &= S_{h}(m,A) + S^{\rm MK}_{\rm liq}(\ph) + 2^{d-1} \ph G(m,A) \ , \\
S_h(m,A) &= \frac{d}{2} (m-1) \ln (2\p A) + \frac{d}{2} (m-1 + \ln m) \ ,  \\
G(m,A) &= d \int_0^\io dr r^{d-1} [q_A(r)^m - \th(r - \s)] \ , \\
q_A(r) &=  \int d\mathbf{ r}' f^{\rm G}_{2A}(\mathbf{ r}') \th( | \mathbf r-\mathbf{ r}' | - \s) =\int_D^\io du \left( \frac{u}{r} \right)^{\frac{d-1}{2}} \frac{ e^{
-\frac{(r-u)^2}{4A} }}{\sqrt{4\p A}} \left[ e^{ -\frac{ru}{2A} } \sqrt{\pi \frac{ru}{A}} I_{ \frac{d-2}{2} } \left( \frac{ru}{2A} \right)\right] \ ,
\end{split}\eeq
where $f^{\rm G}_{A}(r)$ is the $d$-dimensional Gaussian cage given in Eq.~\eqref{eq:AGaussian} and $I_n(x)$ is the modified Bessel function. The last expression for $q_A(r)$
is obtained using bipolar coordinates to compute the convolution~\cite{PZ10}. Remarkably, in odd dimensions the integral over $u$ can be computed analytically, which facilitates the
numerical evaluation of the replicated entropy.

From Eq.~\eqref{eq:SSrep}, we can derive the equation for $A$ from the condition $\partial\SS/\partial A=0$, which reads \beq\label{eq:FA} 1 =  \frac{2^d \f}{d} \frac{A}{1-m}
\frac{\partial G(m,A)}{\partial A} \equiv \frac{2^d \f}{d}  F(m,A) \ . \eeq
The cage radius in the liquid can be obtained by solving this equation in the limit $m\to 1$, and $\bar\D= 2 d A$.
From Eq.~\eqref{eq:FA}, one
sees that the dynamical transition $\f_{\rm d}$ corresponds to the point where $\frac{2^d \f}{d}  \max_A F(1,A) = 1$. Note that for $\f > \f_{\rm d}$ Eq.~\eqref{eq:FA} admits two solutions, but only the
smaller of the two is a stable physical solution~\cite{PZ10}.

\subsubsection{Calculation of the cage size distribution: the cavity method}
\label{sec:cavity}

More information on the distribution of individual cage shapes and sizes can be obtained from the cavity method~\cite{MPV87,MM09}. Its application to the MK model has been developed in
Ref.~\cite{MPTZ11}, where the cavity equations are derived and discussed. Here, we only present the main steps.

\paragraph{\bf Cavity fields and replica symmetric cavity equations --}
In the cavity approach, the system is described by a set of cavity fields $\psi(\mathbf{r})$. Each cavity field describes the probability of finding a particle at position
$\mathbf{r}$, when it is added to a system of $N-1$ particles. The replica symmetric cavity equations provide a recurrence equation for determining these cavity fields
\beq\begin{split} \label{eq:rscav_shift} \psi_0(\mathbf{r}_0) &= \frac1{z_0} \prod_{j=1}^{N_0} \left[ \int d\mathbf{r}_j \psi_{j}(\mathbf{r}_j)
\chi(\mathbf{r}_0 -\mathbf{r}_j+\mathbf{\Lambda}_{0j}) \right] \ , \\
z_0 &= \int d\mathbf{r}_0  \prod_{j=1}^{N_0} \left[ \int d\mathbf{r}_j \psi_{j}(\mathbf{r}_j) \chi(\mathbf{r}_0 -\mathbf{r}_j+ \mathbf{\Lambda}_{0j})\right]\ .
\end{split}\eeq
In this recurrence, the new particle interacts with the $N_0$ other particles, each described by its own cavity field $\psi_{j}(\mathbf{r}_j)$. The interaction is given by the hard-core
constraint $\chi(\mathbf{r}) = e^{-\beta U(r)} =
\theta(r-\sigma)$. %with $\sigma$ the particle diameter.
In this equation
the quenched random variables $\mathbf{\Lambda}_{0j}$ are the random shifts that appear in the Hamiltonian, but they should be independently extracted
at each cavity iteration. They
are independently distributed in the whole volume $V$ with a uniform distribution $P(\mathbf{\Lambda}_{0j}) = 1/V$.
Note that in Ref.~\cite{MPTZ11}, the cavity equations were obtained for a model defined on a random graph that is locally tree-like, corresponding to a situation where $N_0$ remains
finite as $N\to \io$. The method, however, is also applicable to the MK model, where $N_0 = N-1$, corresponding to the fully connected graph~\cite{MPV87}. A convenient way to
obtain the fully connected graph is to first take the limit $N\to\io$ and then $N_0\to\io$. One can show that this procedure is
indeed equivalent to considering $N_0 = N-1$~\cite{MPV87}.

\paragraph{\bf Translational invariance and irrelevance of the random shifts --}
In order to describe the liquid and the glassy states of the MK model, we are interested in solutions of the cavity equation that have statistical translational invariance. To be more
precise, the liquid phase is described by uniform fields $\psi(\mathbf{r})=1/V$ for all particles. Physically, this situation corresponds to particles diffusing everywhere
within the system volume, which mathematically reproduces the virial expansion~\cite{MPTZ11}. In the glass phase, each individual cavity field has the form
$\psi(\mathbf{r}) = f_A(\br - \mathbf R)$, where $f_A(\br)$ is a cage function localized around $\br=0$. The cavity field is thus localized around point $\mathbf R$, but the
localization centers $\mathbf R$ themselves must be uniformly distributed in the whole volume because the glass is globally translationally invariant. Hence, in
Eq.~\eqref{eq:rscav_shift}, when neighbors are picked at random, they are localized around uniformly distributed random positions in space, which makes the random shifts redundant. In
the following, we can thus neglect the random shifts and write the replica symmetric cavity equations as \beq\begin{split} \label{eq:rscav} \psi_0(\mathbf{r}_0) &= \frac1{z_0}
\prod_{j=1}^{N_0} \left[ \int d\mathbf{r}_j \psi_{j}(\mathbf{r}_j)
\chi(\mathbf{r}_0 -\mathbf{r}_j) \right] \ , \\
z_0 &= \int d\mathbf{r}_0  \prod_{j=1}^{N_0} \left[ \int d\mathbf{r}_j \psi_{j}(\mathbf{r}_j) \chi(\mathbf{r}_0 -\mathbf{r}_j )\right]\ ,
\end{split}\eeq
and take $N_0 \to \io$.

\paragraph{\bf The glass phase and the 1RSB cavity equations --}
In the glass phase, as mentioned above, $\psi(\br)$ are random variables described by a probability distribution $Q[\psi]$. In the regime that is here of interest, the glass
is described by the 1RSB cavity equations derived in Refs.~\cite{MPV87,MM09}. From these equations, we obtain that the probability distribution $Q[\psi]$ satisfies the self-consistent
equation \beq \begin{split}
Q[\psi_0] &= \frac{1}{\ZZ_0(m)} \int \prod_{j=1}^{N_0} dQ[\psi_{j}] \, z_0^m \, \d\big[ \text{Eq.~\eqref{eq:rscav}} \big] \ , \\
\ZZ_0(m) &=  \int \prod_{j=1}^{N_0} dQ[\psi_{j}] \, z_0^m \ , \\
 \end{split} \eeq
which makes explicit that the $N_0$ cavity fields describing the neighborhood of the new particle are extracted independently from $Q[\psi]$. The new cavity field is constructed
according to Eq.~\eqref{eq:rscav} and weighted according to $z_0^m$. Note that $z_0$ is the free volume associated with the new particle, and therefore, by varying the free parameter
$m$, one can select glassy states according to their free volume or, equivalently, their internal entropy.

\paragraph{\bf Reconstruction equations --}
Beyond the dynamical transition, ergodicity is broken in the liquid phase, which corresponds to the liquid splitting into many distinct glassy states. It is well known, however, that if configurations are
sampled with the equilibrium Gibbs-Boltzmann measure, then the entropy and pressure are analytic around $\f_\mathrm{d}$ and the equilibrium glass phase is the analytical continuation
of the liquid phase. In order to weight glassy states according to the equilibrium Gibbs-Boltzmann measure, one has to weight them proportionally to their free volume, hence one must set
$m=1$~\cite{MM09}. In the case $m = 1$, the 1RSB equations greatly simplify thanks to a mapping onto the reconstruction formalism~\cite{MM06}. Reconstruction is then done by introducing new
fields $R_\mathbf{r}[\psi(\mathbf{r}')] \equiv \psi(\mathbf{r}) Q [\psi(\mathbf{r}')]$. This change of variable ensures that only the fields that are
localized around point $\br$ contribute to $R_\mathbf{r}[\psi]$.
%Introducing this change of variable in the 1RSB cavity equations and
Using the global translational invariance $R_\mathbf{r}[\psi(\mathbf{r}')] = R_\mathbf{0}[\psi(\mathbf{r}'-\mathbf{r})]$, we conveniently get \beq\label{eq:rec_invp2}
\begin{split}
R_{0}[\psi(\mathbf{r}_0)] & = \int \prod_{j=1}^{N_0} \left[ \frac{ d\mathbf{r}_j \chi( \mathbf{r}_j) } { \int d\mathbf{r}'  \chi(  \mathbf{r}' )} \right] \int dR_0[\psi_{j}(
\mathbf{r}'_j)] \, \d\big( \star \big)
 \ , \\
\star & \hskip10pt \leftarrow \hskip10pt \psi(\mathbf{r}_0) - \frac1{z_0} \prod_{j=1}^{N_0} \left[ \int d \mathbf{r}'_j \psi_{j}(\mathbf{r}'_j) \chi(\mathbf{r}_0  - \mathbf{r}_j -
\mathbf{r}'_j)  \right] \ .
\end{split}\eeq
Note that the reweighting term $z_0^m$ has now disappeared from the equations. Note also that only $R_0[\psi]$ enters the equations and therefore all cavity fields are localized around
the origin. The $\br_j$ in Eq.~\eqref{eq:rec_invp2} are random shifts of the cavity fields that are constrained to be outside a sphere of radius $\s$ around the origin. The neighbors
$j$ are thus localized outside that sphere, which guarantees that around the origin there exists a void to accommodate an additional particle.

\paragraph{\bf Ansatz on the cage shape --}
As discussed in Ref.~\cite{MPTZ11}, numerically solving the cavity equations in Eq.~\eqref{eq:rec_invp2} remains a formidable task. Here, we make a simple ansatz on the cage shape to
facilitate this computation. We first assume that the cavity fields all have the form $\psi_j(\br) = f_{A_j}(\br - \mathbf R)$, where $f$ is a fixed (spherically symmetric) cage shape.
We then choose
either a Gaussian (Eq.~\eqref{eq:AGaussian}) %$f^G_A(\mathbf{r}) = \exp[-\mathbf{r}^2/(2A)]/\sqrt{2 \pi A}^d$
or a ball (Eq.~\eqref{eq:Aball}) cage shape, with $\D_i$ given by Eqs.~\eqref{MSDGaussian} and \eqref{MSDBall}, respectively.
%$f^B_A(\mathbf{r}) = \th( A - \mathbf{r} ^2)/V_d(\sqrt{A})$.
We assume that the cage sizes are distributed according to a function $P_f(A)$ while the centers $\mathbf R$ are uniformly distributed within the volume, as discussed above. We therefore
obtain the ansatz \beq\label{plantingansatz}
\begin{split}
Q[\psi(\mathbf{r})] &= \int dP_f(A) \int \frac{d\mathbf{R}}V \d[\psi(\mathbf{r}) - f_A(\mathbf{r}- \mathbf{R})] \ ,  \\
R_0[\psi(\mathbf{r})] &= \psi(0) Q[\psi(\mathbf{r})] =  \int dP_f(A) \int \frac{d\mathbf{R}}V f_A( \mathbf{R}) \d[\psi(\mathbf{r}) - f_A(\mathbf{r}- \mathbf{R})] \ .
\end{split}\eeq
The above equations show that fields contributing to $R_0[\psi]$ are localized around a point $\mathbf R$ that is distributed according to $f_A(\br)$, and hence $\mathbf R$ is
itself localized close to the origin. Plugging this ansatz in Eq.~\eqref{eq:rec_invp2}, we obtain \beq\label{eq:rec_inv_gamma}
\begin{split}
R_{0}[\psi(\mathbf{r}_0)] & = \int \prod_{j=1}^{N_0} \left[ \frac{ d\mathbf{r}_j \chi( \mathbf{r}_j) } { \int d\mathbf{r}'  \chi(  \mathbf{r}' )} \right] \int dP_f(A_j)
\frac{d\mathbf{R}_j}V f_{A_j}(\mathbf{R}_j) \, \, \d\left[ \psi(\mathbf{r}_0) - \frac1{z_0} \prod_{j=1}^{N_0} q_{A_j/2}[\mathbf{r}_0 -(\mathbf{r}_j + \mathbf{R}_j) ] \right] \ ,
\end{split}\eeq
where \beq q_A(\mathbf{r}) = \int d\mathbf{r}' f_{2A}(\mathbf{r}') \chi(\mathbf{r}-\mathbf{r}') \  \eeq Note that the factor of 2 is introduced to follow the notational convention of Ref.~\cite{PZ10}.

\paragraph{\bf Reconstruction procedure --}
The physical interpretation of the reconstruction equation is quite straightforward. In order to construct a new cavity around the origin, one should draw at random $N_0\to\io$
particles that are located at random positions $\br_j$ outside a sphere of radius $\s$ around the origin. These particles are themselves within a cage, whose size $A_j$ is extracted
from $P_f(A)$. The point $\br_j$ is {\it not} the center of the cage, but a point that is typical of the distribution inside the cage. The cage center is therefore at $\br_j + \mathbf
R_j$, where the shift $\mathbf R_j$ is extracted from the cage shape $f_{A_j}(\mathbf{R}_j)$. Each neighbor rattles around its cage center at $\br_j + \mathbf R_j$, producing an
effective potential that convolutes the HS constraint with the cage shape, $e^{-\beta v^j_{\rm eff}(\mathbf{r}_0)} = q_{A_j/2}[\mathbf{r}_0 -(\mathbf{r}_j + \mathbf{R}_j) ]$. The new cavity
field is then given by the (normalized) exponential of the sum of all effective potentials, $\psi(\mathbf{r}_0)\propto \prod_j q_{A_j/2}[\mathbf{r}_0 -(\mathbf{r}_j + \mathbf{R}_j) ] =
\exp[ -\b \sum_j v^j_{\rm eff}(\mathbf{r}_0) ]$. Finally, we note that although the number of neighbors $N_0$ should be sent to infinity, distant neighbors do not affect the new
cavity field because $q_{A_j/2}(\br)$ tends to 1 when $\br\to\io$. We can therefore introduce an arbitrary spatial cutoff and only consider the neighbors (whose number distribution is Poissonian) that are within this cutoff, and then increase the cutoff until the results converge. This approach is expressed by the following recursive procedure for self-consistently determining
the distribution $P_f(A)$, which is the only remaining unknown in the cavity reconstruction. Note that once $P_f(A)$ has been calculated, one can easily obtain
the distribution of mean square displacements in the cage, $P_f(\D)$, according to Eq.~\eqref{MSDGaussian} or \eqref{MSDBall}. This observable is also easily measured in numerical simulations (and experiments).

\vskip10pt \noindent \noindent\rule{8cm}{0.4pt}
%------------------------------------------------------------------------------------------------------------------------------------------------------
\\
{\bf Procedure-Reconstruction-MK}
\begin{enumerate}
\item Consider a spherical shell $\s < r < \s + \s_{\rm cut}$ of volume $V_0$ (the upper bound $\s_{\rm cut}$ should be sufficiently large for the results to be independent of it).
Consider a number of centers $N_0$ distributed according to a Poisson law with average $\overline{N_0} = \r V_0$. Uniformly draw these sphere centers $\mathbf{r}_j$ within the shell.

\item Independently draw $N_0$ cage radii $A_j$ from $P_f(A)$, and $N_0$ displacements $\mathbf{R}_j$ from $f_{A_j}(\mathbf{R}_j)$.

\item From these $N_0$ random variables, derive a new cavity field \beq\label{newfield_MK} \psi(\mathbf{r}_0) = \frac{\prod_{j=1}^{N_0} q_{A_j/2}(\mathbf{r}_0 -(\mathbf{r}_j +
\mathbf{R}_j) )} {\int d\mathbf{r}'_0 \prod_{j=1}^{N_0} q_{A_j/2}(\mathbf{r}'_0 -(\mathbf{r}_j + \mathbf{R}_j) )} \ . \eeq

\item  Compute the mean square displacement in the new cavity as \beq\begin{split}
\langle \br_0 \rangle &= \int d\mathbf{r}_0 \, \mathbf{r}_0 \, \psi(\mathbf{r}_0) \\
 \langle \d \mathbf{r}^2_0 \rangle &= \int d\mathbf{r}_0 \, (\mathbf{r}_0-\langle \mathbf{r}_0 \rangle)^2 \, \psi(\mathbf{r}_0) = \D_{\rm
new}/2 \ . \label{eq:Anew}
\end{split}\eeq
The value $\D_{\rm new}$ is the long-time mean square displacement corresponding to Eq.~\eqref{eq:Ai}.
It allows one to determine the new cage parameter $A_{\rm new}$ that enters in the new cage
shape in Eq.~\eqref{eq:AGaussian} (or \eqref{eq:Aball}), according to Eq.~\eqref{MSDGaussian} (or \eqref{MSDBall}).

\item Repeat steps (1-4) to get $\NN$ samples $A_{\rm new}$ in order to construct a new distribution $P_f(A_{\rm new})$.

\item Repeat (5) until the distribution converges $P_f(A_{\rm new}) \simeq P_f(A)$ within the statistical error.

\item From the convergent $P_f(A)$ compute the distribution of mean square displacements, $P_f(\D)$, using Eq.~\eqref{MSDGaussian} (or \eqref{MSDBall}).
\end{enumerate}
\noindent\rule{8cm}{0.4pt}
%------------------------------------------------------------------------------------------------------------------------------------------------------
\\

\paragraph{\bf Numerical details --}
In principle, the above procedure provides a theoretical way to compute $P_f(A)$, but practically it must be implemented numerically, with two additional tricks.

First, we note that it is difficult to calculate the normalization of the cavity field  $\psi(\mathbf{r}_0)$ in Eq.~\eqref{newfield_MK}, because one has to integrate
over the whole space. It is more convenient to compute the variance $\langle \d \mathbf{r}^2_0
\rangle$ in Eq.~\eqref{eq:Anew} using the Metropolis algorithm without explicitly obtaining  $\psi(\mathbf{r}_0)$. We can then write Eq.~\eqref{newfield_MK} as \beq \psi(\mathbf{r}_0)
=  \frac{\tilde{\psi}(\br_0)}{\int d\br_0 \tilde{\psi}(\br_0)}, \eeq where the non-normalized probability $\tilde{\psi}(\br_0)$ is \beq \tilde{\psi}(\br_0) \equiv \prod_{j=1}^{N_0}
q_{A_j/2}[\mathbf{r}_0 -(\mathbf{r}_j + \mathbf{R}_j) ]. \eeq From this expression, it is clear that $\tilde{\psi}(\br_0)$ is analogous to the Boltzmann factor in the Gibbs measure
with an effective potential $\mathcal{H}_{\rm eff}(\br_0)$, $\tilde{\psi}(\br_0) = e^{-\beta\mathcal{H}_{\rm eff}(\br_0)}$. We can thus use the standard Monte Carlo (MC) algorithm to sample
any average quantity, such as $\langle \d \mathbf{r}^2_0 \rangle$, with acceptance rate \beq {\rm acc}(\br_0^{\rm old} \to \br_0^{\rm new}) = \min \{1, \tilde{\psi}(\br_0^{\rm
new})/\tilde{\psi}(\br_0^{\rm old}) \}. \eeq
Interestingly, we actually derived from the cavity formalism a ``local'' MC simulation. %(Note that we only have a single variable $\br_0$ compared with $N$ variables $\br_1, \ldots \br_N$ in a normal MC simulation of the entire system.)
In this local MC sampling, the positions of all the particles, except for the caged particle at $\br_0$,  are fixed and their vibrational contribution to the motion of the caged
particle is integrated into the effective potential $\mathcal{H}_{\rm eff}(\br_0)$. In our simulations, we perform $4\times10^5$ MC steps with step size $0.1\sqrt{\bar{A}}$ to calculate each cage size.

%{\color{red} define standard MC: number of steps and size of steps.}

Second, we have to remove hopping from the cavity procedure, or otherwise the cavity solution does not properly converge (see Fig.~\ref{fig:cavity}). To achieve this task, during the
calculation of $A_{\rm new}$ in the local MC simulations, we record the spatial trajectory of $\br_0$. We then check if any hopping occurs during this trajectory using
the detection algorithm described in Sec.~\ref{sec:hopping_detection}.  We only include $A_{\rm new}$ in the statistics of $P_f(A_{\rm new})$ if no hopping is detected, as otherwise the
particle is not truly caged. Once hopping is removed, our results indicate that the cavity solution properly converges when $\ph > \ph_{\rm d}$ (see Fig.~\ref{fig:cavity}). We represent the distribution $P_f(A)$ by a number $\NN = 10^4 \gg \overline{N_0}$ of samples $A_k$, $k=1\cdots \NN$.
% At each iteration step we extract
%the cages of the neighbors by taking at random $N_0$ of the $\NN$ samples, and we iterate the procedure above to construct $\NN$ new samples that constitute the
%new population. We repeat this until we see that the distribution $P(A)$ is stabilized and does not change under additional iterations within the statistical error.

\begin{figure}[ht]
\centerline{\includegraphics [width = 2.5in] {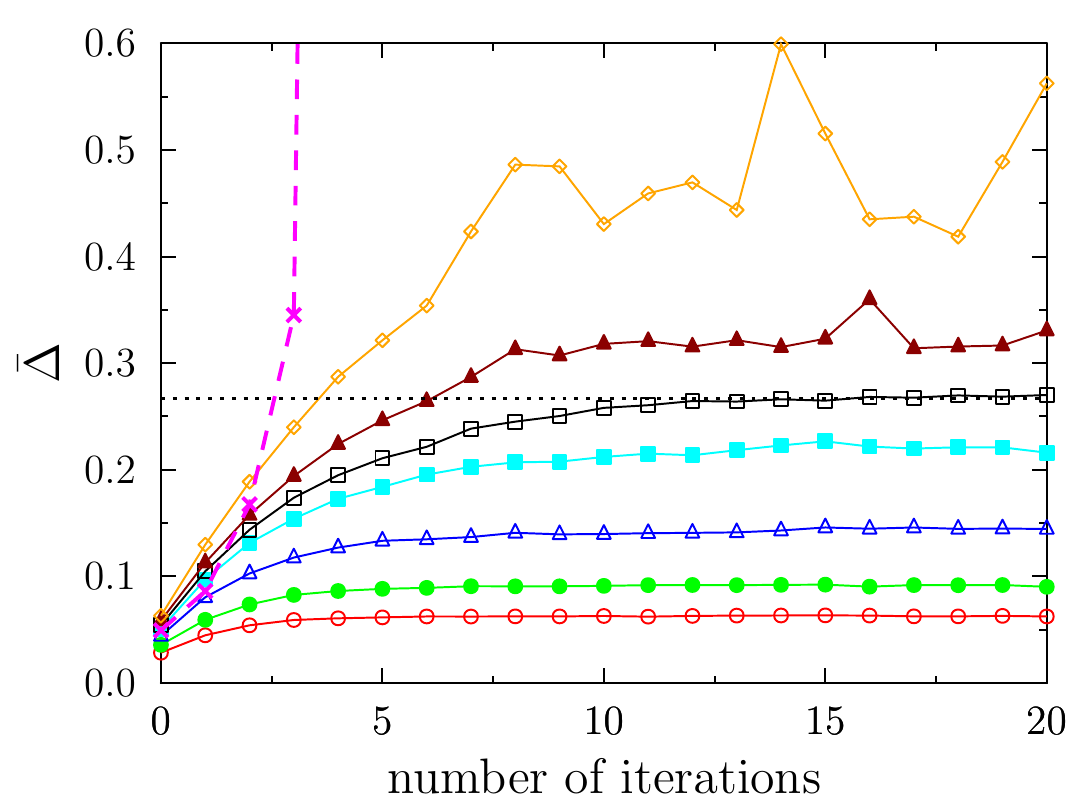} } \caption{Evolution of mean square displacement $\bar{\D}$ under iteration of
the cavity reconstruction,
at (solid lines, from bottom to top)
$\varphi = 2.50, 2.20, 1.95, 1.80, 1.75, 1.70, 1.60$ in $d=3$. The solution becomes completely unstable above $\D_d = 0.267$ (dotted black line),
as predicted by the replica method. If
hopping is not removed, the solution diverges quickly when $\varphi$ approaches $\varphi_{\rm d}$. See, for instance, the unfiltered results for $\varphi = 1.95$ (pink dashed line).} \label{fig:cavity}
\end{figure}

\subsubsection{Comparing theoretical predictions with simulations}
\label{sec:phid}
We first show that the mean square displacement $\bar{\D}$ predicted from both the replica and the cavity methods is generally in good agreement with the simulation data (Fig.~\ref{fig:A}). The simulation $\bar{\D}$ is extracted from the asymptotic time limit of the MSD data, according to the MCT scaling (see Eq.~\eqref{eq:above} below).
Close to $\ph_{\rm d}$, the replica theory predicts a scaling (Fig.~\ref{fig:A})
\beq
|\bar\D(\ph) - \D_{\rm d}| \sim |\ph - \ph_{\rm d}|^{1/2} \label{eq:scalingA}\ ,
\eeq
where $\D_{\rm d} = \bar\D(\ph_{\rm d})$,
that is consistent with the MCT prediction. However, around $\ph_{\rm d}$ precisely determining $\bar{\D}$ from either simulations or cavity reconstruction requires a careful consideration of hopping. The simulation and the cavity data therefore unsurprisingly deviate from Eq.~\eqref{eq:scalingA} in that regime (Fig.~\ref{fig:A}). %For the same reason, we are only able to unambiguously determine $\ph_{\rm d}$ from the cavity method where hopping is pre-excluded in the theoretical construction.

Cavity reconstruction provides a theoretical prediction for the distribution $P_f(\D)$ of individual mean square displacements.
In order to obtain individual cages from simulation, we use Eq.~\eqref{eq:Ai}
at $t = 2$, which is sufficiently long for the cages to form, but not so long that a large fraction of particles have hopped. Note that at densities well above $\ph_{\rm d}$, hopping is so rare that
this choice of timescale is irrelevant. As discussed in the main text, our theoretical results agree well with simulations, and are independent of the Gaussian or of the ball ansatz for the shape
functional. The replica calculation for $\bar{\D}$ using the Gaussian functions and $P_f(\D) = \delta(\D - \bar{\D})$ also agrees with the simulation results (Fig.~\ref{fig:A}).

\begin{figure}[ht]
\centerline{\includegraphics [width = 4in] {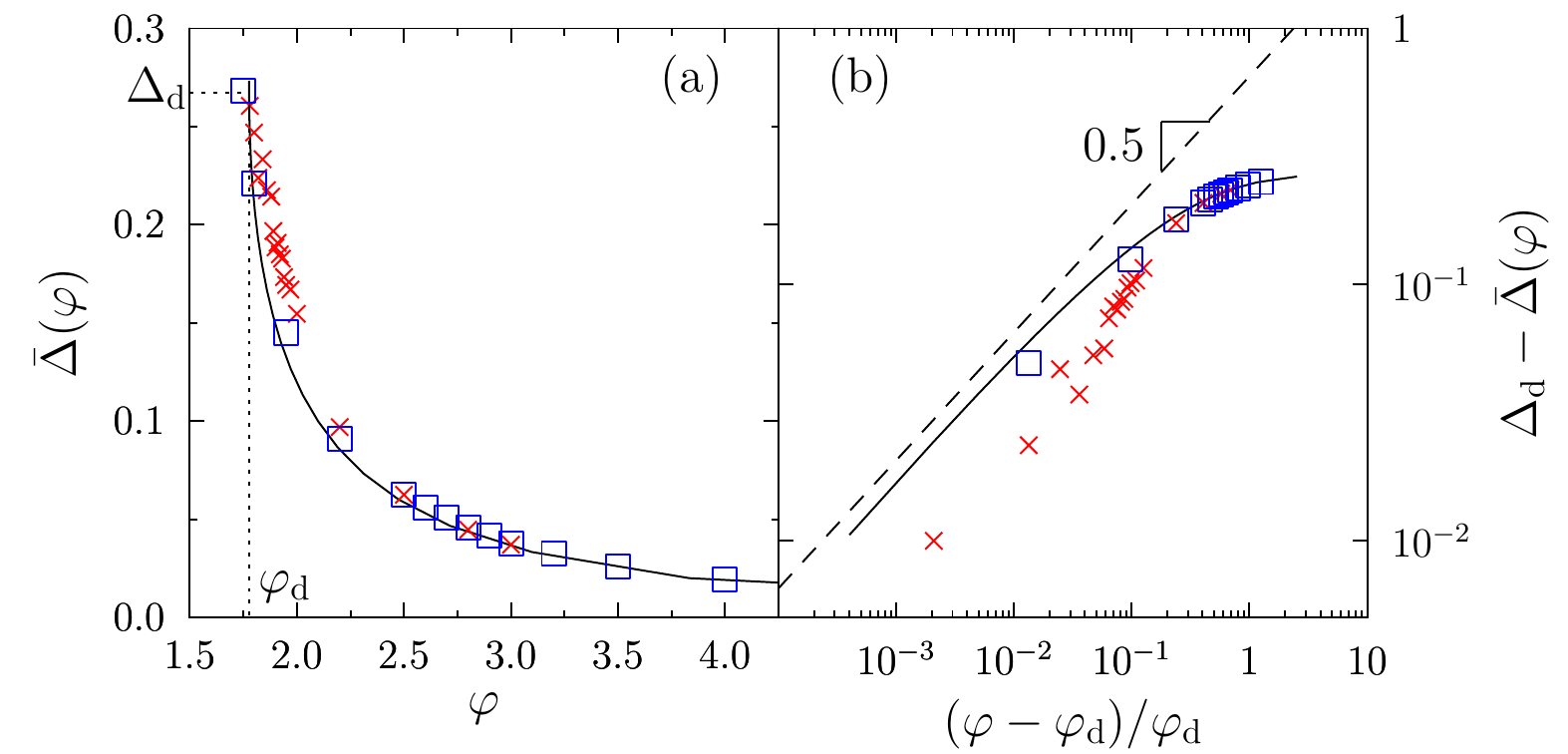} } \caption{(a) The $d=3$ mean square displacement
$\bar{\D}(\ph)$ obtained from replica method (black line), cavity reconstruction (blue
squares), and MD simulations (red crosses). The replica result for $\ph_{\rm d}$ correspond to the point where the theoretical replica line has a square root singularity.
(b) The theoretical results are consistent with the scaling form in Eq.~(\ref{eq:scalingA}), but deviations are observed in the MD data close to $\varphi_\mathrm{d}$, due to the ambiguity in determining cage sizes when hopping is significant.} \label{fig:A}
\end{figure}

%{\color{blue}  
In summary, we find a basic consistency between our MD simulations and theoretical calculations, including (i) the replica calculation with a Gaussian anzatz for the cage shape and a $\delta$-function approximation for the cage size distribution function $P_f(\Delta) \approx \delta(\Delta - \bar{\Delta})$, and (ii) the cavity method with both Gaussian and ball anzatzs. It has been shown that in the limit $d \to \infty$, the theoretical result (of replica calculation) is independent of the cage shape anzats \cite{KPZ12}, and we also expect it to be independent of the method we use (replica/cavity). In finite dimensions, weak dependence is expected, but according to our results presented here, it is insignificant compared to the numerical accuracy of
the resolution of the cavity equations.
%}

\subsection{Caging dynamics: mode-coupling theory (MCT) and beyond}
In this section, we compare the MD results with the dynamical caging behavior predicted by MCT. The MCT scalings are found to only be consistent with our data when $\ph <
\ph_{\rm SER}$. Above $\ph_{\rm SER}$, MCT predictions are violated, which is well captured by the breakdown of SER and is a consequence of entangling caging with hopping, as
discussed in Sec.~\ref{sec:hopping}. It is important to note  that we here only refer to MCT as the general scaling laws predicted by the schematic MCT equation~\cite{Go09},
which can be
also independently derived from the static framework~\cite{Parisi2013}. The traditional MCT kernel being incorrect for the MK model~\cite{MK11}, the numerical
MCT predictions are indeed unsuitable for comparison.

\subsubsection{Testing the mode-coupling theory}
\label{sec:MCT} We first compile the MCT predictions tested in our study. The derivations of these predictions as well as many important physical interpretations can be found
in Ref.~\cite{Go09} and references therein.  We denote $\epsilon = \frac{\ph-\ph_{\rm d}}{\ph_{\rm d}}$ as the distance from the dynamical transition, and $\tau_\epsilon$ as the
characteristic time for the $\beta$-relaxation. Note that MCT does not predict any breakdown of the SER, so we do not distinguish between the $\alpha$-relaxation time $\tau_\alpha$ and the
diffusion time $\tau_D$ in this analysis ($\tau_D \sim \tau_\alpha \gg \tau_\epsilon$). Below the dynamical transition $\ph < \ph_{\rm d}$, MCT predicts that the time evolution of
the MSD has the form
\begin{equation}
\Delta_{-}(t)  =
\begin{cases}
\D_d - B|\epsilon|^{1/2}\left(\frac{t}{\tau_\epsilon}\right)^{-a}, & t \ll \tau_\epsilon,\\
\D_d + C\left(\frac{t}{\tau_D}\right)^b + \frac{t}{\tau_D}, & t \gg \tau_\epsilon,
\end{cases}
\label{eq:below}
\end{equation}
where $B$ and $C$ are density-independent constants, and the exponents $a$ and $b$ are related by the exponent parameter  $\lambda$ as
\begin{equation}
\lambda = \frac{[\Gamma(1-a)]^2}{\Gamma(1-2a)} = \frac{[\Gamma(1+b)]^2}{\Gamma(1+2b)} \label{eq:lambda}.
\end{equation}

Equation~\eqref{eq:below} shows that the relaxation process of $\D$ can be divided into three regimes: (i) an early $\beta$-relaxation towards the plateau $\D_d$, $\D_d - \D_{-}(t) \sim
t^{-a}$, (ii) a late $\beta$-relaxation leaving from the plateau, $\D_{-}(t) \sim t^b$, and (iii) a diffusive process that is linear in time $\D_{-}(t) \sim t$. We stress that, as described in Sect.~\ref{sec:MSD}, before regime (i), there is a ballistic regime characterized by a microscopic time that is much smaller than $\tau_\epsilon$ and is not included in
Eq.~\eqref{eq:below}.
%Second, the $\alpha$-relaxation is usually well approximated by a stretched exponential law (Kohlrausch-Williams-Watts function) $\D \sim
%e^{-(t/\tau_\alpha)^\beta}$, but this process may be too fast to be observed in the MSD. Following the convention in Ref.~\cite{KA95a}, we thus simply merge the late
%$\beta$-relaxation with the diffusive regime in Eq.~\eqref{eq:below}.

One of the most important predictions made by MCT is that, upon approaching $\ph_{\rm d}$, a power-law divergence should be observed for
\beq \tau_D \sim |\ph-\ph_{\rm d}|^{-\gamma},
\label{eq:tauD} \eeq and \beq \tau_\epsilon \sim |\ph-\ph_{\rm d}|^{-1/2a}, \eeq where the exponents are related via
\begin{equation}
\gamma = \frac{1}{2a}+\frac{1}{2b}. \label{eq:gamma}
\end{equation}

Beyond $\ph_{\rm d}$, MCT then predicts
\begin{equation}
\Delta_{+}(t) =
\begin{cases}
 \D_{\rm d} - B|\epsilon|^{1/2}\left(\frac{t}{\tau_\epsilon}\right)^{-a}, & t \ll \tau_\epsilon \ ,\\
 \D_{\rm d} \ , & t \gg \tau_\epsilon \ .
\end{cases}
\label{eq:above}
\end{equation}
These scalings are tested by the following procedure.

\vskip10pt \noindent \noindent\rule{8cm}{0.4pt}
%------------------------------------------------------------------------------------------------------------------------------------------------------
\\
{\bf Procedure-Testing-MCT}
\begin{enumerate}
\item Obtain $\ph_{\rm d}$ and $\Delta_{\rm d}$ from the replica calculation.
\item Fit $\tau_D$ according to Eq.~\eqref{eq:tauD} (see Fig.~1 of main paper) with the theoretical $\ph_{\rm d}$, in order to obtain the exponent $\gamma$. A consistent power-law scaling is only observed below some density $\ph_{\rm SER}$; above $\ph_{\rm SER}$, $\tau_D$ becomes smaller than
the MCT predictions, implying that an additional relaxation process starts to interfere with the dynamics. This observation suggests that when we fit the diffusivity data,
only the data below $\ph_{\rm SER}$ should be used. If instead we treat $\ph_{\rm d}$ as a fitting parameter for the entire density range, then we end up with shifted
values $\tilde{\ph}_{\rm d}$ and $\tilde{\gamma}$ \beq \tau_D \sim |\ph-\tilde{\ph_{\rm d}}|^{-\tilde{\gamma}}, \label{eq:tauD2} \eeq From the analysis presented in the main text, it
is clear that MCT actually fails when $\ph >  \ph_{\rm SER}$, and thus the apparent power-law fitting of Eq.~\eqref{eq:tauD2} is not reliable.  Results for $\ph_{\rm d}$, $\tilde{\ph}_{\rm d}$, $\gamma$ and $\tilde{\gamma}$ can be found in Table~\ref{table:phi}.
%(Similar results for HS are in Tables~\ref{table:phiHS} and ~\ref{table:gammaHS}).
\item Determine $a$ and $b$ (Table~\ref{table:phi}) from $\gamma$ using Eqs.~\eqref{eq:lambda} and \eqref{eq:gamma}.
\item Test the dynamical behavior of $\D(t)$ (Eq.~\eqref{eq:below}) below $\ph_{\rm d}$, and determine the constants $B$ and $C$. Note that here we have fixed all the other parameters, $\varphi_{\rm d}, \Delta_{\rm d}, a$ and $b$ from previous steps. Figure~\ref{fig:MCT}(a) and (b) show that when $\ph < \ph_{\rm SER}$, Eq.~\eqref{eq:below} is satisfied in the entire time regime; when $\ph < \ph_{\rm SER}$, it is only satisfied in the early $\beta$-relaxation regime.
\item Using the same constant $B$, check the MCT dynamics above $\ph_{\rm d}$ using Eq.~\eqref{eq:above}. When $\ph$ is not too far away from $\ph_{\rm d}$, $\D(t)$ does not strictly saturate to a plateau as predicted by MCT, and the scaling behavior of the intermediate time regime is modified.
\end{enumerate}
\noindent\rule{8cm}{0.4pt}

%\begin{table}
%\begin{ruledtabular}
%\begin{tabular}{c | c c c c c }
%  $d$  & $\gamma$ &  $\tilde{\gamma}$ &$a$ & $b$ & $\l$ \\
%  \hline
%2 & 4.59(4) & 5.77(4) & 0.19 & 0.26 & 0.92 \\
%3 & 3.27(7) & 4.95(4) & 0.25 & 0.40 & 0.85 \\
%4 & 2.9(1) & 4.50(7)  & 0.28 & 0.46 & 0.80 \\
%5 & 2.67(8) & 4.04(6) & 0.29 & 0.52 & 0.78 \\
%6 & 2.65(8) & 3.75(4) & 0.30 & 0.53 & 0.76 \\
%\end{tabular}
%\end{ruledtabular}
%\caption{Numerical results for the MCT exponents for the MK model in $d = 2 - 6$.
%} \label{Tatable:gammaMK}
%\end{table}

%\begin{table}
%\begin{ruledtabular}
%\begin{tabular}{c | c c c c c }
%  $d$  & $\gamma$ &  $\tilde{\gamma}$ & $a$ & $b$ & $\l$ \\
%  \hline
%3   &   1.72(3) & 2.8(1) & 0.40 & 1.05 & 0.47 \\
%4   &   1.92(3) & 2.26(4) & 0.37  & 0.86  & 0.57  \\
%5   &   1.95(3) & 2.23(6) & 0.37 &   0.84  & 0.58   \\
%6   &   2.00(3) & 2.22(6) & 0.36 & 0.80  & 0.60   \\
%7   &   2.0(1)  & 2.23 (7) & 0.36 & 0.80 & 0.60  \\
%8   &   2.15(5) & 2.15(5) & 0.34 &  0.71 & 0.65  \\
%\end{tabular}
%\end{ruledtabular}
%\caption{Numerical results for the MCT exponents for HS in $d = 3 - 8$.
%} \label{table:gammaHS}
%\end{table}

\begin{figure}[ht]
\centerline{\includegraphics [width = 5.5in] {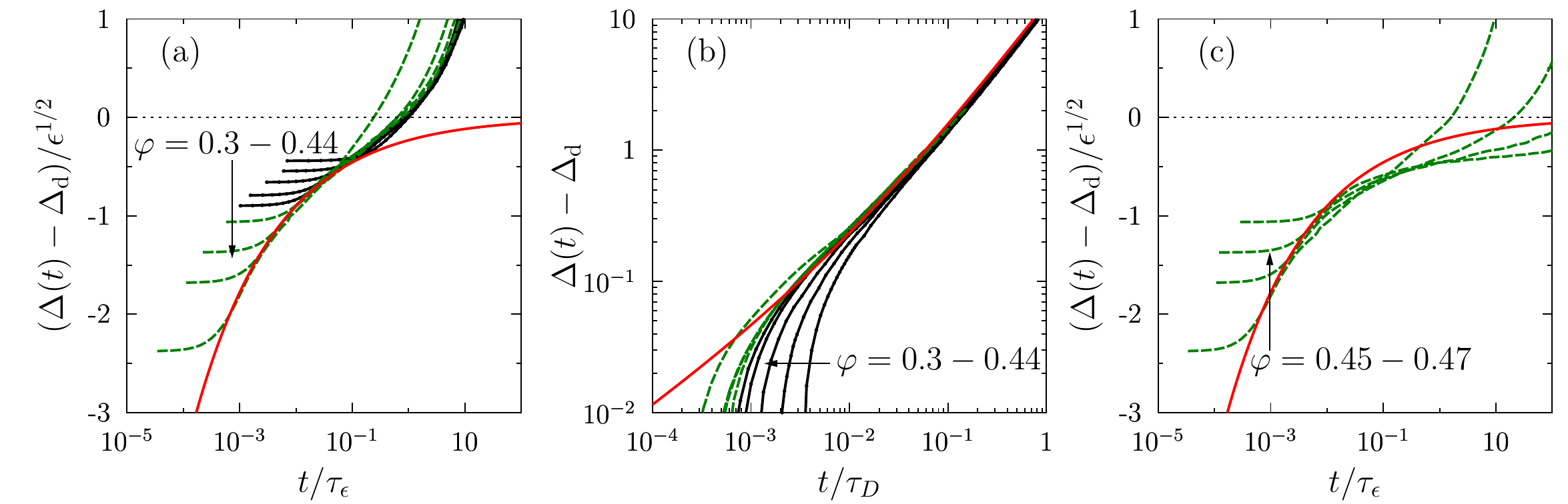} } \caption{Testing MCT scalings for the MK model in $d=6$. Below $\varphi_{\rm d}$, the MD data for $\varphi = 0.30, 0.35, 0.38, 0.40, 0.41, 0.42, 0.43, 0.435, 0.44$ are fitted to Eq.~(\ref{eq:below}) (red lines) for (a) the early
$\beta$-relaxation, and (b) the late $\beta$-relaxation together with diffusion, using fitting parameters $B=0.073$ and $C=1.3$. (c) Above $\varphi_{\rm d}$, the MD data $\varphi = 0.45, 0.455, 0.46, 0.47$ are compared to
the early $\beta$-relaxation scaling in Eq.~(\ref{eq:above}) with the same value of $B$. A good agreement is found for the entire time regime when $\varphi < \varphi_{\rm SER}$ (black
solid lines). When $\varphi > \varphi_{\rm SER}$ (green dashed lines), we only observe a good agreement for the early $\beta$-relaxation regime, which suggests that at later times hopping mixes with the MCT dynamics. } \label{fig:MCT}
\end{figure}

\subsubsection{Breakdown of the Stokes-Einstein relation}

The above analysis shows that the MCT scalings start to break down close to the dynamical transition $\ph_{\rm d}$. To further investigate this property, we look at the scaling
relation between the diffusion time $\tau_D$ and the relaxation time $\tau_\alpha$. If SER were obeyed, we should obtain $\tau_D \sim \tau_\alpha$. As shown in Fig.~1 in the main paper,
SER breaks down when $\ph > \ph_{\rm SER}$ as
\begin{equation}
\tau_D \sim \tau_\alpha^{1-\omega},
\end{equation}
where the exponent $\omega = 0.22$ is invariant with $d$.

The breakdown of SER is beyond the MCT description. Interestingly, we observe that three phenomena happen at the same density $\ph_{\rm SER}$: (i) violation of MCT scalings, (ii) violation of SER, and (iii) the hopping characteristic time $\tau_{\rm h}$ becoming comparable with $\tau_D$. Our interpretation of these observations is presented in the main text. %: In the classical mean-field framework (in both dynamic and static senses), approaching to $\ph_{\rm d}$, a particle starts to be confined in a cage within the relaxation time scale. However, in reality a particle is not always well caged and it has a certain probability to `hop' outside of the cage. In this case, two cages are connected by a narrow channel, and the particle is traveling from one cage to another through the channel. Below $\ph_{\rm SER}$, the dynamics is fast enough and it is impossible to distinguish between the MCT-like relaxation and the hopping dynamics. In this regime MCT is valid. Above $\ph_{\rm d}$, the MCT-like relaxation becomes so slow that hopping also contributes to the dynamics. The two processes are mixed and the dynamics becomes really complex. The single time scale scenario of the relaxation and diffusion does not hold anymore -- the SER breaks down.

\subsection{Percolation of the cage network}
Because cages can be connected via hopping channels, it is natural to examine how the cages are topologically connected. We find that the network of cages spans the system below
density $\ph_{\rm p}$ ($\ph_{\rm p} > \ph_{\rm d}$). Above $\ph_{\rm p}$, only local cage clusters are formed and particles become strictly confined. We show that this phenomenon can
be mapped onto a void percolation transition, which belongs to the same universality class as regular percolation.

\subsubsection{Mapping the glass transition to a void percolation transition}
To do the mapping, we first consider the simplest case, where we assume that all the neighbors of a given particle are frozen, $P_f(A) = \delta(A)$. We want to know if the caged particle
can move to another cage without overlapping with other particles. Equivalently, we can rescale the size of neighbors as $\sigma \to
2\sigma$, and look for a hopping path for the point representing the caged particle in the leftover void space (see Fig.~\ref{fig:percolation}a).

We next consider the situation where cage sizes are not zero. In this case, particle $j$ is rattling inside a cage with radius $\sqrt{A_j}$, whose distribution is a density dependent
function $P_f(A)$.  If a certain channel were closed in the first case, there is now a possibility for it to be open because the particles bounding that channel are now thermally
moving. Because we are interested in the upper bound for percolation, i.e., the best case scenario for hopping, we rescale particle sizes as (see Fig.~\ref{fig:percolation}a):
\beq \sigma \to 2(\sigma - 2 \sqrt{A_j}),
\label{eq:percolation_mapping} \eeq where ${A_j}$ is drawn from $P_f(A)$. If no path in void space is found by this construction, then the particle is confined. Strictly
speaking, this procedure only works for cage shapes with sharp boundaries, like the ball function in Eq.~\eqref{eq:Aball}. For a Gaussian cage in Eq.~\eqref{eq:AGaussian}, the confined
particle always has a finite (but vanishingly small) probability to hop, even if the cage is found to be closed in the percolation mapping. In the following percolation analysis, to avoid any possible confusion, we assume that all cages have ball shapes, which corresponds to
assuming that this probability tail is negligible.

%Our strategy for the percolation analysis is: (i) first calculate $P(A)$ at density $\ph$ from the cavity method, (ii) plant random configurations at the same density (as in the cavity
%method, see Fig.~3 in the main paper) and map the configurations to systems of polydisperse spheres using Eq.~\eqref{eq:percolation_mapping}, and (iii) check if the void space in the
%mapped configurations percolates. The above procedure is repeated at different $\ph$, until we find the percolation threshold $\ph_{\rm p}$, which is an upper bound for the dynamical
%glass transition $\ph_{\rm d}$. The methodological details for step (iii) are given below.

\begin{figure}[ht]
\centerline{ \includegraphics [width = 3.5in] {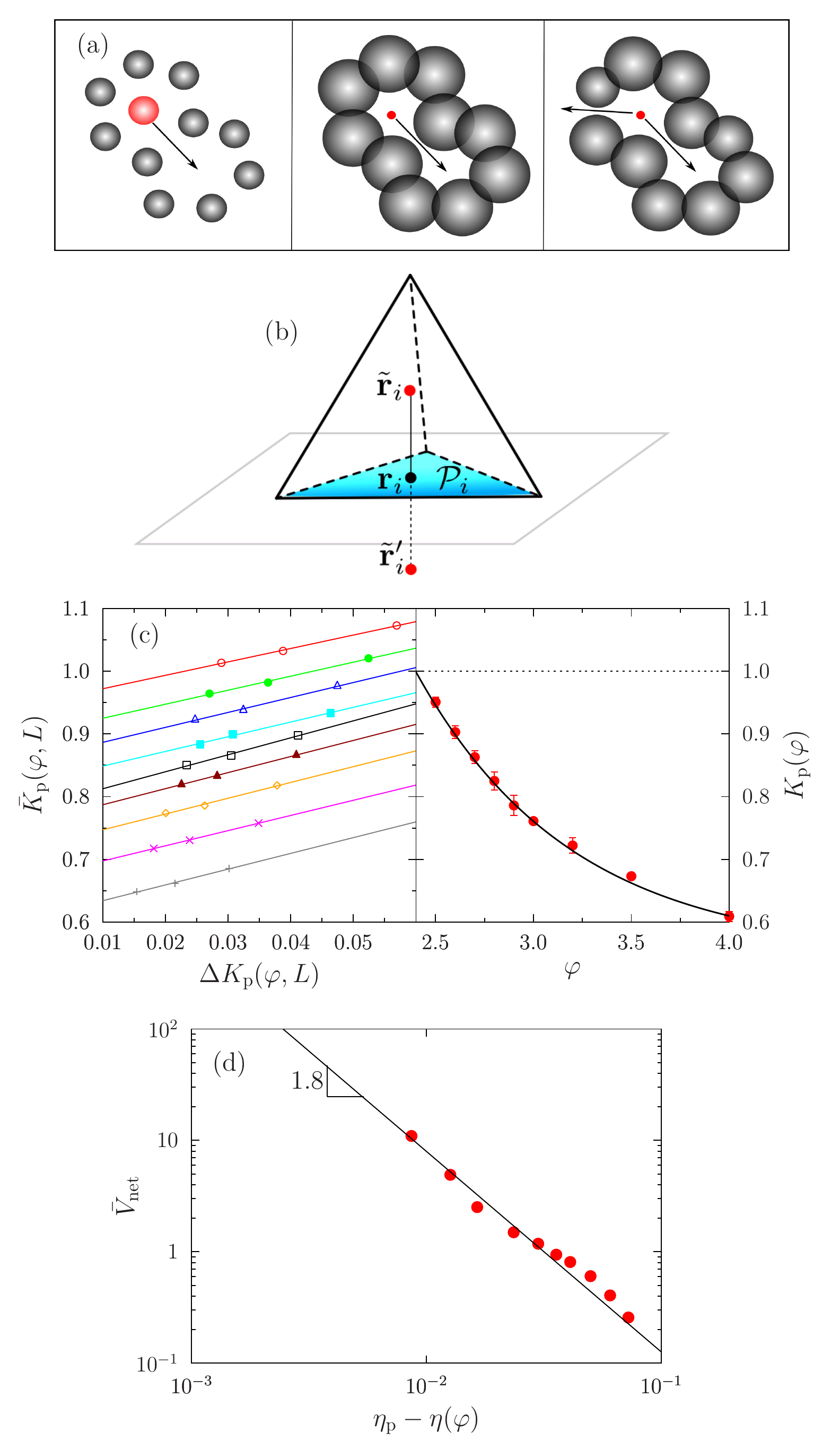} } \caption{Percolation analysis. (a) Mapping the glass transition problem to a void percolation transition according to the rescaling law $\sigma \to 2 \sigma$ (from left to middle panels), and according to $\sigma \to 2(\sigma - 2\sqrt{A_j})$ (from left to right panels). In the latter case, additional hopping paths may be found. (b) Calculating the radical Voronoi cell $\mathcal{P}_i$ (blue triangle) of sphere $i$ in dimension $d$ ($d=2$ in this example). The problem is mapped onto a Voronoi tesselation in dimension $d+1$, where the tetrahedron is the Voronoi cell of $\tilde{\br}_i$ in the mapped $d+1$ configuration. (c) Determining $\ph_{\rm p}$ in $d=3$. Left: estimation of the percolation threshold $K_{\rm p}(\varphi)$ from the finite-size analysis
of Eq.~(\ref{eq:finite_size}). Right: determining the percolation density $\ph_{\rm p} = 2.4$ from $K_{\rm p}(\ph_{\rm p}) = 1$. The black line is an exponential fit of the data points. (d) Scaling of the mean cluster volume is consistent with Eq.~(\ref{eq:mcv}) using the exponent $\gamma_{\rm p} = 1.8$ given by
standard lattice percolation.} \label{fig:percolation}
\end{figure}

\subsubsection{Methodology for determining the void percolation threshold}
{\bf Mapping the void space onto a network via the Voronoi tessellation --} For void percolation, unlike for lattice percolation, or for continuous percolation (which is the dual problem to void percolation),
the pre-defined network is not trivial to extract. It is has been shown, however, that the void space between monodisperse spheres can be represented by a network obtained by
Voronoi tessellation~\cite{Kerstein1983}. This method can also be generalized to polydisperese systems via radical Voronoi tessellation~\cite{vanderMarck1996}. In the network
representation, nodes are Voronoi vertices, and links are the edges of the Voronoi polyhedra. If any link passes through one or more sphere, then it is blocked and should be removed from
the network. After this network is constructed, we check if there exists a percolated path from the center of the network to the system boundary.

{\bf Algorithm for radical Voronoi tessellation of polydisperse spheres in any $d$--} Because our systems have a range of cage sizes, i.e., they map onto spheres with a polydisperse diameters, we develop a method to produce the radical Voronoi
tessellation for a given configuration.
%~\cite{Rycroft}. 
The basic idea is to map the radical Voronoi tessellation in dimension $d$ to a Voronoi tessellation in dimension $d+1$, and then
use Qhull~\cite{Barber1996} to compute the Voronoi tessellation.

For the standard Voronoi tessellation, the Voronoi cell for sphere $i$ consists of space points $\br$ that satisfy the relation
\begin{equation}
  \label{eq:voro1}
  |\br -\br_i| < |\br - \br_j|,
\end{equation}
for any $j\neq i$. The radical Voronoi tessellation is a generalization of this definition for unequal sized spheres:
\begin{equation}
  \label{eq:voro2}
  |\br-\br_i|^2 - R_i^2 < |\br-\br_j|^2 - R_j^2,
\end{equation}
where $R =\sigma/2$ is the particle radius.

In order to map the radical Voronoi tessellation to a Voronoi tessellation, we denote $R_{\rm max}$ the maximum radius, and introduce a set of points in dimension $d+1$,
\smash{$\tilde{\br}_i=(\br_i^{1},\br_i^{2},\ldots,\br_i^{d},\sqrt{R_{\rm max}-R_i^2})$}, where $i = 1 \ldots N$.  The first $d$ coordinates of $\tilde{\br}_i$ are the same as $\br_i$,
and the final coordinate is a function of the sphere radius. We further introduce a set of dual points \smash{$\tilde{\br}'_i=(\br_i^{1},\br_i^{2},\ldots,\br_i^{d}, -\sqrt{R_{\rm
max}-R_i^2})$}, as images of $\tilde{\br}_i$'s with respect to the last coordinates. For each pair $\{\tilde{\br}_i, \tilde{\br}'_i\}$, we find the $d$-dimensional polygon $\mathcal{P}_i$
that is the common Voronoi boundary between these two points (Fig.~\ref{fig:percolation}b). According to the definition in Eq.~\eqref{eq:voro1}, it is clear that any point $\tilde{\br}$ in
$\mathcal{P}_i$ can be written as $\tilde{\br} = (\br, 0)$, and $\tilde{\br}$ satisfies
\begin{equation}
  \label{eq:voro3}
  |\tilde{\br} -\tilde{\br}_i| < |\tilde{\br} - \tilde{\br}_j|,
\end{equation}
which is equivalent to
\begin{equation}
  \label{eq:voro5}
  |\br - \br_i|^2 + R_{\rm max}^2 -R_i^2 < |\br - \br_j|^2 + R_{\rm max}^2 - R_j^2.
\end{equation}
Because this relation is exactly the definition of the radical Voronoi cell in Eq.~\eqref{eq:voro2}, we have proven that the $d$-dimensional polygon $\mathcal{P}_i$ is  the radical Voronoi cell of
sphere $i$ in the original configuration.

%\begin{figure}[ht]
%\centerline{ \includegraphics [width = 2.7in] {voronoi_method} } \caption{Calculating the radical Voronoi cell $\mathcal{P}_i$ (blue triangle) of sphere $i$ in dimension $d$ ($d=2$ in this example). The problem is mapped to a Voronoi %tesselation in dimension $d+1$, where the tetrahedron is the Voronoi cell of $\tilde{\br}_i$ in the mapped $d+1$ configuration.}
%\label{fig:voronoi_method}
%\end{figure}

{\bf Determining the percolation threshold from the scaling theory--} The percolation threshold can be determined by finite-size scaling~\cite{Rintoul1997}. Note that in the void
percolation analysis, the variable of interest is the volume fraction of void space $\eta$, and not directly the volume fraction $\ph$ \cite{vanderMarck1996,Elam1984}. Because our
planted configuration is essentially a Poisson process of overlapping spheres, however, we have
\beq
 \eta (\ph) = \left[1 - \frac{V_d(\sigma/2)}{V} \right]^{N} \approx e^{-\frac{N V_d(\sigma/2)}{V}} = e^{-\ph}.
\eeq
%where $V_d(\sigma/2)$ is the volume of a sphere.

Let $\eta_{\rm p}= e^{-\ph_{\rm p}}$ be the percolation threshold in the infinite system-size limit. For a system of finite linear size $L \sim V^{1/d}$, the average effective
percolation threshold $\bar{\eta}_{\rm p}(L)$ and its variance  $\Delta \eta_{\rm p}(L)$  are linearly related~\cite{Stauffer1994}
\begin{equation}
|\bar{\eta}_{\rm p} (L) - \eta_{\rm p}| \sim \Delta \eta_{\rm p}(L),
\end{equation}
which allows one to numerically determine $\eta_{\rm p}$. Because the polydispersity associated with the cage distribution $P_f(A)$ varies with density $\ph$, we cannot, however,
directly use this relation. At each $\ph$,  we instead modify the rescaling rule of Eq.~\eqref{eq:percolation_mapping} by adding a factor $K(\ph, L)$ \beq \sigma \to
2(\sigma - 2 \sqrt{A_j}) K(\ph, L), \label{eq:percolation_mapping2} \eeq and use a binary search to find the percolation threshold $K_{\rm p}(\ph, L)$ for each configuration. We then
calculate $\bar{K}_{\rm p}(\ph, L)$ and  $\Delta K_{\rm p}(\ph, L)$ over 1000 independent realizations, and use a similar relation \beq |\bar{K}_{\rm p} (\ph, L) - K_{\rm p}(\ph)| \sim
\Delta K_{\rm p}(\ph, L), \label{eq:finite_size} \eeq
to determine $K_{\rm p}(\ph)$ (Fig.~\ref{fig:percolation}c). We finally compute $\ph_{\rm p}$, such that  $K_{\rm p}(\ph_{\rm p}) = 1$. Our system is thus percolated at
$\ph_{\rm p}$, without the extra rescaling factor $K_{\rm p}(\ph)$ (Fig.~\ref{fig:percolation}c).

%\begin{figure}[ht]
%\centerline{ \includegraphics [width = 4in] {percolation_threshold} } \caption{(a) Estimation of the percolation threshold $K_{\rm p}(\varphi)$ from the finite-size analysis
%Eq.~(\ref{eq:finite_size}) in 3d. (b) Determining the percolation density $\ph_{\rm p} = 2.4$ from $K_{\rm p}(\ph_{\rm p}) = 1$. The black line is an exponential fit of the data points.} \label{fig:percolation_threshold}
%\end{figure}

To check the universality of the percolation transition, we examine the scaling of the mean cluster size $\bar {V}_{\rm net}$, where $V_{\rm net}$ is the total volume of the cluster of
cages connected to the planted central cage. According to percolation theory, $\bar {V}_{\rm net}$ diverges at the percolation threshold as a power-law with exponent $\gamma_{\rm p}$:
\begin{equation}
\bar {V}_{\rm net} \sim |\eta (\ph) - \eta_p|^{-\gamma_p}. \label{eq:mcv}
\end{equation}
Figure~\ref{fig:percolation}d shows that our results are in agreement with $\gamma_{\rm p} = 1.8$~\cite{Stauffer1994} given by lattice percolation, in support of the two problems sharing a same universality class.

%\begin{figure}[ht]
%\centerline{\includegraphics [width = 2.5in] {mcv}} \caption{Scaling of the mean cluter volume is consistent with Eq.~(\ref{eq:mcv}) with exponent $\gamma_{\rm p} = 1.8$ given by
%standard lattice percolation.} \label{fig:mcs}
%\end{figure}

{\bf Procedure for determining $\ph_{\rm p}$--} Based on the above discussion, we summarize the procedure for determining $\ph_{\rm p}$. \vskip10pt \noindent
\noindent\rule{8cm}{0.4pt}
%------------------------------------------------------------------------------------------------------------------------------------------------------
\\
{\bf Procedure-$\ph_{\rm p}$-determination}
\begin{enumerate}
\item For a given density $\ph$, obtain the distribution $P_f(A)$ from the cavity method.
\item Plant a configuration with linear system size $L$ such that the central particle is compatible with all neighbors (this requirement is the same as for the cavity method, see Sec.~\ref{sec:cavity}).
\item Rescale the particle sizes following Eq.~(\ref{eq:percolation_mapping2}).
\item Find the percolation threshold $K_{\rm p}(\ph, L)$ for the configuration using a binary search. To determine if the void space is percolated:
\begin{itemize}
\item Map the $d$-dimensional configuration of rescaled polydisperse spheres to a $(d+1)$-dimensional configuration of monodisperse spheres.
\item Use a Voronoi protocol to calculate the Voronoi tesselation of the $(d+1)$-dimensional configuration.
\item Map back the $(d+1)$-dimensional Voronoi tesselation to the $d$-dimensional radical Voronoi tesselation.
\item From the radical Voronoi tesselation, construct a network.
\item Determine if the network is percolated.
\end{itemize}
\item Repeat (1-4) to get $\bar{K}_{\rm p}(\ph, L)$ and  $\Delta K_{\rm p}(\ph, L)$.
\item Vary $L$ and repeat (5) to get $\bar{K}_{\rm p}(\ph, L)$ and  $\Delta K_{\rm p}(\ph, L)$ at different $L$, and use the finite scaling Eq.~(\ref{eq:finite_size}) to obtain $K_{\rm p}(\ph)$.
\item Vary $\ph$ and repeat (6) to get $K_{\rm p}(\ph)$ at different $\ph$, and find $\ph_{\rm P}$ such at $K_{\rm p}(\ph_{\rm p}) = 1$.
\end{enumerate}
\noindent\rule{8cm}{0.4pt}

\section{Hopping}
\label{sec:hopping}

In this section, we detail how we detect hopping events in numerical simulations, and describe the hopping dynamics of the MK model at a phenomenological level. Theoretical investigations are left to future study.

\subsection{Detecting hopping}
\label{sec:hopping_detection} We follow the algorithm of Refs.~\cite{CWKDBHR10,CDB09} to detect hopping events in both the MD simulations and the numerical evaluation of the cavity
equations. Below we briefly summarize the procedure.

\vskip10pt \noindent \noindent\rule{8cm}{0.4pt}
%------------------------------------------------------------------------------------------------------------------------------------------------------
\\
{\bf Procedure-detection-hopping}
\begin{enumerate}
\item Run simulations and save particle trajectories.
\item Determine the cage size $\Delta_i$ of each particle as discussed in Sec.~\ref{sec:MSD}. (We relax the assumption of Refs.~\cite{CWKDBHR10,CDB09} that all cages have the same size.)
\item Split each single-particle trajectory $X(0<t<t_{\rm tot})$ into two subsets $X_1(0<t_1<t^*)$ and
$X_2(t^*<t_2<t_{\rm tot})$, and measure the mean square distance between the two sub-trajectories
\begin{equation}
\delta(t^*) = \xi(t_c)[\langle d_1(t_2)^2 \rangle_{t_2} \langle d_2(t_1)^2 \rangle_{t_1}]^{1/2},
\end{equation}
where $d_j(t_k)$ is the distance between the point at time $t_k$ and the center of mass of the subset $X_j$ ($j,k = 1,2$), and $\xi(t_c) = \sqrt{\frac{t^*}{t_{\rm
tot}}\left(1-\frac{t^*}{t_{\rm tot}}\right)}$ is a normalization factor. Find the time $t^*_{\rm max}$ such that $\delta(t^*_{\rm max})$ is maximum.

\item For particle $i$, if $\delta_i(t^*_{\rm max})
>  \Delta_i$, hopping is detected, and the process is repeated recursively for each sub-trajectory until $\delta_i(t^*_{\rm max}) <  \Delta_i$ in each sub-trajectory.
\end{enumerate}
\noindent\rule{8cm}{0.4pt}
\\

Following this procedure, we save a sequence of hopping times Note that in this study we are only interested in the time of the first hopping, which is equivalent to the time during
which the particle is trapped in a cage before escaping. Because facilitation is reduced in the MK model, especially  at high $\varphi$, we do not specifically distinguish between the first and the subsequent hopping
events. The algorithm generally works well at densities $\ph > \ph_{\rm d}$, as shown in Fig.~\ref{figure:hopping_detect}. Close to $\ph_{\rm d}$, however, hopping is mixed with other
relaxation processes, and cages are not clearly defined. Detecting hopping indeed then becomes more sensitive to the specific cutoff thresholds.

\begin{figure}[ht]
\centerline{ \includegraphics [width = 4in] {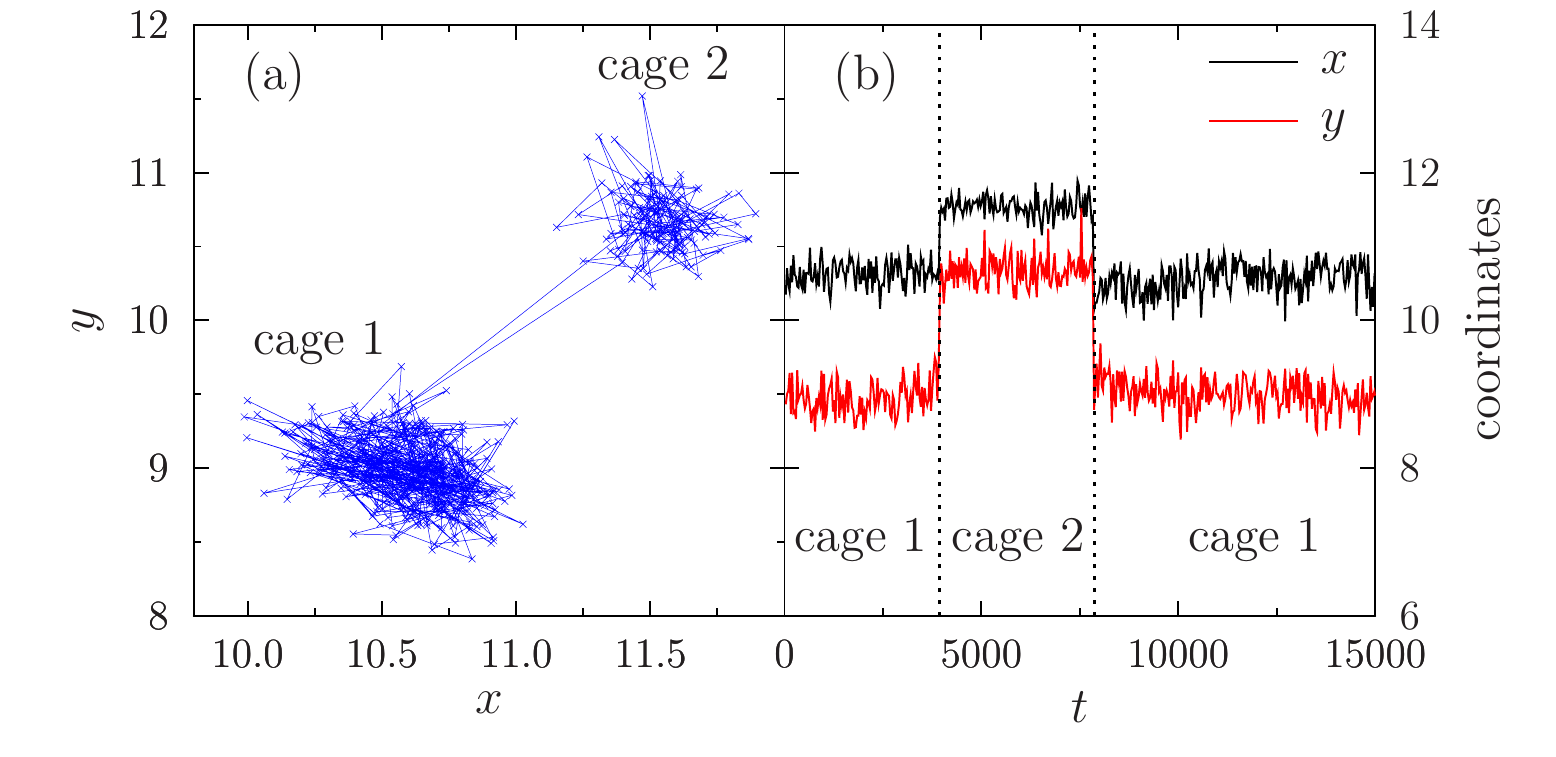} } \caption{An example of hopping detection in $d = 2$  at $\ph = 2.40$. (a) The particle trajectory clearly
reveals two well formed cages. (b) The hopping between cages are visualized in the time series, with the two detected hopping times (dotted lines) at $t=3948.0$
and $t=7863.6$. } \label{figure:hopping_detect}
\end{figure}

\subsection{Hopping dynamics}
Empirically, we find that the above detected hopping time $t$ follows a power-law distribution (see Fig.~3 of the main paper)
\begin{equation}
p_{\rm h}(t) \sim t^{-\mu}, \label{eq:hopping_dist}
\end{equation}
with exponent $\mu < 1$. We can write its cumulative distribution function as
\begin{equation}
G_{\rm h} (t) = (t/\tau_{\rm h})^{1-\mu},
\end{equation}
where  $\tau_{\rm h}$ is the characteristic hopping time scale, representing the time needed for all particles to hop, $G_{\rm h}(\tau_{\rm h}) = 1$. As shown in Fig.~3 of the main paper, both $\mu$ and $\tau_{\rm h}$ depend on $\ph$. In particular, $\tau_{\rm h}$ is roughly an exponential function of $\ph$
\begin{equation}
\tau_{\rm h} \sim e^{\alpha \ph},
\end{equation}
which suggests that there is no diverging density for $\tau_{\rm h}$.

\bibliography{HS,glass,hopping,hopping_SI}
%\begin{thebibliography}{10}
\end{widetext}

\end{document}